\documentclass[aps,prd,twocolumn,amssymb,superscriptaddress,nofootinbib,eqsecnum]{revtex4}
\pdfoutput=1
\usepackage{amsmath}
\usepackage{graphicx}
\usepackage{amssymb}
\usepackage{esint}

\makeatletter

\@ifundefined{textcolor}{}
{%
 \definecolor{BLACK}{gray}{0}
 \definecolor{WHITE}{gray}{1}
 \definecolor{RED}{rgb}{1,0,0}
 \definecolor{GREEN}{rgb}{0,1,0}
 \definecolor{BLUE}{rgb}{0,0,1}
 \definecolor{CYAN}{cmyk}{1,0,0,0}
 \definecolor{MAGENTA}{cmyk}{0,1,0,0}
 \definecolor{YELLOW}{cmyk}{0,0,1,0}
 }

\begin{document}

\title{Luminosity distance in Swiss cheese cosmology with randomized voids. II. Magnification probability distributions}
%
%
\newcount\hh
\newcount\mm
\mm=\time
\hh=\time
\divide\hh by 60
\divide\mm by 60
\multiply\mm by 60
\mm=-\mm
\advance\mm by \time
\def\hhmm{\number\hh:\ifnum\mm<10{}0\fi\number\mm}
\def\dofigures{1}


\author{\'{E}anna \'{E}. Flanagan}

\email{flanagan@astro.cornell.edu}

\address{Laboratory for Elementary Particle Physics, Cornell University, Ithaca,
NY 14853, USA}

\address{Center for Radiophysics and Space Research, Cornell University, Ithaca,
NY 14853, USA}

\author{Naresh Kumar}

\email{nk236@cornell.edu}

\address{Laboratory for Elementary Particle Physics, Cornell University, Ithaca,
NY 14853, USA}

\author{Ira Wasserman}

\email{ira@astro.cornell.edu}

\address{Laboratory for Elementary Particle Physics, Cornell University, Ithaca,
NY 14853, USA}

\address{Center for Radiophysics and Space Research, Cornell University, Ithaca,
NY 14853, USA}

\author{R. Ali Vanderveld}

\email{rav@kicp.uchicago.edu}

\address{Kavli Institute for Cosmological Physics, University of Chicago,
Chicago, Illinois 60637, USA}

\begin{abstract}

We study the fluctuations in luminosity distances due to gravitational
lensing by large scale ($\gtrsim35$ Mpc) structures, specifically
voids and sheets. We use a simplified {}``Swiss cheese'' model consisting
of a $\Lambda$CDM Friedman-Robertson-Walker background in which a
number of randomly distributed non-overlapping spherical regions are
replaced by mass compensating comoving voids, each with a uniform
density interior and a thin shell of matter on the surface. We compute
the distribution of magnitude shifts using a variant of the method
of Holz \& Wald (1998), which includes the effect of lensing shear. The standard
deviation of this distribution is $\sim0.027$ magnitudes and the
mean is $\sim0.003$ magnitudes for voids of radius 35 Mpc, sources
at redshift $z_s=1.0$, with the voids chosen so that 90\% of the mass
is on the shell today.  The standard deviation
varies from $0.005$ to $0.06$ magnitudes as we vary the
void size, source redshift, and fraction of mass on the shells today.
If the shell walls are given a finite thickness of $\sim 1$ Mpc, the standard deviation
is reduced to $\sim 0.013$ magnitudes.
This standard deviation due to voids is a factor
$\sim 3 $ smaller than that due to galaxy scale structures.
We summarize our results in terms of a fitting formula that is accurate to $\sim 20\%$, and also build a simplified analytic model
that reproduces our results to within $\sim 30\%$.
Our model also allows us to explore the domain of validity of weak
lensing theory for voids.  We find that for 35 Mpc voids, corrections
to the dispersion due to lens-lens coupling are of order $\sim 4\%$, and corrections to due shear
are $\sim 3\%$.
Finally, we estimate the bias due to source-lens clustering in
our model to be negligible.
\end{abstract}
\maketitle

\def\be{\begin{equation}}
\def\ee{\end{equation}}
\def\bea{\begin{eqnarray}}
\def\eea{\end{eqnarray}}
\def\nn{\nonumber}
\newcommand{\bes}{\begin{subequations}}
\newcommand{\ees}{\end{subequations}}
\def\bec{\begin{centering}}
\def\eec{\end{centering}}
\def\bftheta{\boldsymbol{\theta}}

\def\ba{\begin{eqnarray}}
\def\ea{\end{eqnarray}}
\def\yvec{{\bf y}}
\def\Avec{{\bf A}}
\def\Mvec{{\bf M}}
\def\bvec{{\mbox{\boldmath $b$}}}
\def\DM{\Delta\Mvec}
\def\Dell{\Delta\ell}
\def\Dphi{\Delta\phi}
\def\delbar{{\overline{\Delta}}}
\def\Cvec{{\bf C}}
\def\delvec{{\mbox{\boldmath $\delta$}}}
\def\dotprod{{\bf\cdot\,}}
\def\sigvec{{\mbox{\boldmath$\sigma$}}}
\def\Sigvec{{\mbox{\boldmath$\Sigma$}}}
\def\Tr{{\rm Tr}}
\def\det{{\rm det}}
\def\thetavec{{\mbox{\boldmath$\theta$}}}
\def\Svec{{\bf S}}
\def\ehat{{\bf\hat e}}
\def\ommi{\Omega_{M,i}}
\def\omm{\Omega_M}
\def\Mshell{M_{\rm shell}}
\def\Mpart{M_{\rm part}}
\def\Rpart{R_{\rm part}}
\def\fshell{f_{\rm shell}}
\def\ain{a_{\rm in}}
\def\hin{h_{\rm in}}
\def\zin{z_{\rm in}}
\def\rin{r_{\rm in}}
\def\xin{x_{\rm in}}
\def\uin{u_{\rm in}}
\def\zetain{\zeta_{\rm in}}
\def\aomm{a_{\Omega_M}}
\def\fnow{f_{\rm now}}
\def\yshell{y_{\rm shell}}
\def\pinterior{p_{\rm void}}
\def\ypart{y_{\rm part}}
\def\ppart{p_{\rm part}}
\def\Npart{N_{\rm part}}
\def\Nbm{\langle N(b)\rangle}
\def\ycosmo{y_{\rm cosmo}}
\def\Mpc{{\rm Mpc}}
\def\kmpspmpc{\,{\rm km\,s^{-1}\Mpc^{-1}}}
\def\Msun{M_\odot}
\def\riso{r_{\rm iso}}
\def\vcirc{v_{\rm circ}}
\def\zhat{{\hat z}}
\def\Rhat{{\hat R}_s}
\def\rmax{R_{\rm max}}
\def\Rmax{\rmax}
\def\rhmax{{\hat R}_{\rm max}}
\def\psimax{{\Psi_{\rm max}}}
\def\Iinf{I_\infty}
\def\ymean{\ypart}
\def\Dm{\Delta m}
\def\gE{\gamma_E}
\def\Neff{N_{\rm eff}}
\def\xtil{{\tilde x}}
\def\qci{q_{i,c}}
\def\bci{b_{i,c}}
\def\Var{{\rm Var}}

\section{Introduction}
\label{sec:intro}

\subsection{Background and Motivation}

A number of surveys are being planned to determine luminosity distances
to various different astronomical sources, and to use them to
constrain properties of the dark energy or modifications to gravity
that drive the cosmic acceleration.
It has long been recognized that perturbations to luminosity distances
from weak gravitational lensing will be a source of error for these
studies, both statistical and systematic
\cite{lensingnoise1,HW,Valageas0,HL,Munshi}.  For supernovae the lensing noise
becomes significant only at high redshifts \cite{SNok}, but for gravitational wave
sources the lensing noise dominates over the intrinsic luminosity
scatter \cite{HH,Ghirlanda}.  Theoretical predictions for the magnification
probability distribution can be folded into the data analysis of
surveys to improve the results \cite{Luca}, and in
particular it is possible to exploit the known non-Gaussian nature of
this distribution \cite{Hirata}.
In addition, it is possible to treat the ``lensing noise'' in
luminosity distances as a signal in its own right, which provides
useful information \cite{lensingsignal}.  (A tentative detection of this
signal in supernovae data has been claimed in Ref.\ \cite{detection}.)
For these reasons, it is useful to have a detailed understanding of the
magnification probability distribution.

There are a number of methods that have been used to study the effects
of weak lensing on luminosity distances:
\begin{itemize}

\item Weak lensing theory
can be used to predict the variance of the magnification distribution
from the matter power spectrum \cite{key-1}.  However,
the accuracy of this approach is limited and in particular it does not
allow one to probe the non-Gaussian tails of the
distribution\footnote{We note however that there is a proposal for an
  approximate ``universal probability distribution'' for
  magnifications that takes
as input only the variance of the distribution as predicted by weak
lensing theory, and which would allow prediction of the non-Gaussian
tails \protect{\cite{lensing2}}.}.

\item One can use numerical ray tracing using the results of
  cosmological simulations of large scale structure, such as the
  Millennium simulation \cite{Millennium}
and the Coyote Universe project \cite{coyote}, see, eg.\ Ref.\ \cite{Li}.
This approach is highly accurate and is based on a realistic density
distribution.  However it requires substantial computational power and is also
limited in some other respects.
The largest simulations to date are
are confined
to a cube of comoving size $z\sim0.16$,
so only a limited range of source redshifts can be considered. Although
the calculations evolve large scale structure nonlinearly, it is impractical
to get a continuous description of the evolution, which is needed
for computing the perturbations to light ray paths; only snapshots
of the density distribution are available. Finally, because the calculations
required to evolve the matter distribution are formidable,
it can be difficult to comprehensively survey the space of the underlying
parameters of the model, such as the primordial perturbation spectrum.

\item A third approach is to use simplified analytical models of the distribution
  of matter that allow rapid computation of the full probability
  distribution of magnifications, see, eg., Refs.\
  \cite{HW,HL,KM09,KM11}.

\end{itemize}

In this paper we follow the third approach.  We
develop an idealized {}``Swiss cheese''
model \cite{VFW,HW,SNISI,KKNNY,BTT07} of large scale structure to
study the effect of density inhomogeneities on luminosity distances.
Our model
is complementary to many of the existing models
in that we focus on lensing
produced by structures at the largest scales, voids and sheets, rather
than that produced by individual galaxies and halos, the focus of many
existing models.

\subsection{Our void model}

In {}``Swiss cheese'' models \cite{VFW,HW,SNISI,KKNNY,BTT07},
the Universe contains a network of spherical, non-overlapping, mass-compensated
voids. The voids are chosen to be mass compensated so that the
potential perturbation vanishes outside each void.
We idealize these models even further by assuming that each
void consists of a central, uniformly underdense region surrounded
by a zero thickness shell. Mass flows outward from the evacuated interior and is then trapped on the wall. Although it would be more
realistic to consider voids with smooth density profiles, this very
simplified model should capture the essence of the effect of large
scale density inhomogeneities on luminosity distances. Since voids
in the observable Universe tend to be surrounded by shells that are
relatively thin compared to the size of their evacuated interiors,
the idealization of zero thickness may not be a severe simplification,
particularly because we expect that the main effect of inhomogeneities
on the luminosity distance depends only on the integral of the density
contrast along the line of sight from the source to the observer.
A key feature of our idealized models is that they can be evolved
in time continuously and very simply.

Within the context of this highly idealized class of models, we study
the distribution of magnitude shifts relative to what would be found
in a smooth cold dark matter (CDM) model of the Universe with a cosmological
constant, $\Lambda$, for different void sizes and present day interior
underdensities, and for a range of different source redshifts. Moreover,
although we shall use a Newtonian description that is valid as long
as the void radii are small enough compared with the Hubble length
$H_{0}^{-1}$, the calculations can be made fully relativistic if
desired. (We discuss some corrections that are higher order in $H_{0}R$, where $R$ is the void radius.)

This paper is a follow-up to our earlier work \cite{VFW} (henceforth
VFW08), in which we considered the effect of a randomized set of voids
with a single and rather large comoving radius, 350 Mpc, using a
particular model for a smooth underdense interior inside a mass compensated
shell. That study found that for a source with redshift $z_s=1.8$, the mean magnitude
shift relative to smooth flat, CDM for an ensemble of realizations
of large scale voids was unimportant (-0.003), but the distribution
of magnitude shifts was fairly broad, with a standard deviation of
about 0.1. Here, we consider a wider range of redshifts and void sizes,
and compute magnitude shifts relative to a more realistic $\Lambda$CDM
background with matter density today $\Omega_M=0.3$ and dark energy density today
$\Omega_{\Lambda}=0.7$.

\subsection{Predictions for lensing noise}

Our results for the standard deviation $\sigma_m$ of the magnitude
shifts are summarized by the approximate fitting formula
\be
\sigma_m
\approx
(0.027 \pm 0.0007)
\left( \frac{R}{35 \, {\rm Mpc}}  \right)^\alpha
\left( \frac{f_0}{0.9} \right)^\beta \left( \frac{z_s}{1.0}
\right)^\gamma.
\label{fit}
\ee
Here $R$ is the comoving radius of the voids, $z_s$ is the source
redshift, and $f_0$ is the fraction of the total void mass in its shell
today.  The exponents are $\alpha = 0.51 \pm 0.03$, $\beta = 1.07
\pm
0.04$, $\gamma = 1.34 \pm 0.05$.
This fit is accurate to $\sim 20 \%$
for $35 \, {\rm Mpc} \, \le R \le 350 \, {\rm Mpc}$, $0.01 \le f_0 \le
0.9$, and $0.5 \le z_s \le 2.1$.
The mean magnitude shift is again unimportant, roughly a factor of ten
smaller than the standard deviation (\ref{fit}).

Our result (\ref{fit}) is computed in the limit of zero shell
thickness.  This idealization is not very realistic, since as we
discuss in Sec.\ \ref{sec:analyticalmodel} below there is a
logarithmic divergence in the variance of the lensing convergence in
the zero thickness limit.  This divergence arises from rays that pass
very near to the void walls.  The variance in the magnitude shift,
however, is finite because of the nonlinear dependence of magnitude
shift on lensing convergence; the divergence is cut off at lensing
convergences of order unity.  (The divergence can also get regulated by
finite sampling effects; see Sec. \ref{sec:analyticalmodel}).
To address this issue we also consider a more realistic model
with void walls of some finite thickness $\Delta r$. We estimate in
Sec.\ \ref{sec:finite} that for $f_0 = 0.9$, $R = 35$ Mpc, and $z_s = 1.0$, the
standard deviation in magnitude shift is
\be
\sigma_m \approx 0.013 \sqrt{ 1 + 0.23 \ln \left( {1 \, {\rm Mpc} \over \Delta r} \right) },
\label{finitewidth}
\ee
a factor of $\sim 2$ smaller than the thin-shell limit (\ref{fit}) for
$\Delta r = 1$ Mpc.

The rms magnitude shift (\ref{finitewidth}) due to voids is a factor of $\sim
3 $ smaller than that computed from individual galaxies and halos
\cite{HL}, in accord with expectations from weak lensing theory using
the power spectrum of density perturbations (see Ref.\ \cite{Munshi}
and Appendix \ref{appA}).  Thus lensing due to voids is subdominant
but not negligible.

We also use our model to estimate the sizes of various nonlinear effects
that go beyond linear, weak-lensing theory.  We estimate that
for $R = 35$ Mpc voids, the
dispersion $\sigma_m$ is altered by $\sim 4\%$ by lens-lens coupling,
by $\sim 3 \%$ by shear.  There are also large nonlinearities ($\sim
30\% - 40\%)$ in our
model that arise from the nonlinearity of void evolution.
These results are qualitatively in agreement
with some previous studies of nonlinear deviations from weak lensing
theory \cite{nonlin1,nonlin2,nonlin3}.

We also study the
source-lens clustering effect \cite{SLC}, the fact
sources are more likely to be located in high density regions,
which enhances the probability of a lens being located near the source.
We estimate that the corresponding bias in the distribution of
magnifications is negligible in our model.

\subsection{Organization of this paper}

This paper is organized as follows. Section \ref{sec:lensingmodel} reviews our Swiss cheese
void model. We discuss how the voids evolve in an FRW background and
describe the model parameters.
Next, we describe how our void locations are randomized, by choosing
impact parameters randomly as light rays exit one void and enter the
next. Finally, we describe our method of computing the magnification.
Section \ref{sec:analyticalmodel} describes our simple analytical
model which reproduces the results of the simulations to within $\sim
30\%$.  It also describes a modification of our void model in which
the shell walls are given a finite thickness, and gives the
corresponding analytical results.
Section \ref{sec:results} gives the results of our Monte Carlo
simulations for the the probability distributions
of magnifications, and discusses the dependence of the variance
on the parameters of the model.
In Section \ref{sec:bias}, we study the source-lens clustering effect
and the associated bias.  Section \ref{sec:conclusions} summarizes our
results
and their implications.  In Appendix \ref{appA} we discuss the
power spectrum of our void model and the corresponding weak lensing
prediction.  Appendix \ref{appB} reviews the derivation of the
method we use to compute the magnification distribution.  Finally,
Appendix \ref{appC} is a comparison of our results with other
recent studies of lensing due to voids \cite{CZ,KM09,KM11,BTT08,Szybka,Biswas}.
Our results are broadly consistent with these previous studies but our
model is simpler in several respects.

\section{Simple Model of Lensing due to Voids}
\label{sec:lensingmodel}

In this section we describe our simplified Swiss cheese model of large
scale voids, and explain how we compute the distribution of
magnifications in the model.

\subsection{Newtonian model of a single void}
\label{sec:voidmodel}

As discussed in the introduction, we
will consider void radii $R$ ranging from
35 Mpc to 350 Mpc, which are small compared to the Hubble length.
Therefore we can use Newtonian gravity to describe each void; the
corresponding error is of order $\left(H_{0}R\right)^{2}\ll1$ which we
ignore.

We choose the background cosmology in which we place our voids to
be an FRW Universe with matter fraction $\Omega_M$ and cosmological
constant fraction $1- \Omega_M$.
We denote by $a_{\rm ex}(t)$ the corresponding scale factor,
which is
normalized so that $a_{\rm ex}(t)=1$ today. It satisfies the
Friedman equation
\begin{equation}
\left(\frac{\dot{a}_{\rm ex}}{a_{\rm
      ex}}\right)^{2}=H_{0}^{2}\left(\frac{\Omega_{M}}{a_{\rm ex}^{3}}+1-\Omega_{M}\right),\label{eq:1}
\end{equation}
where $H_{0}$ is the Hubble parameter,
which
has the solution
\begin{equation}
\frac{3H_{0}t\sqrt{1-\Omega_{M}}}{2}=\sinh^{-1}\left(\frac{a_{\rm ex}^{3/2}}{a_{\Lambda}^{3/2}}\right).
\label{eq:2}
\end{equation}
Here $a_{\Lambda}=\left(\Omega_{M}/\left(1-\Omega_{M}\right)\right)^{\frac{1}{3}}$
is the scale factor at which the cosmological constant starts to
dominate.

Our void model consists of a spherical region of constant comoving
radius $R$, with a uniform density interior surrounded by
a thin shell.  We assume that the void is mass compensated, so the total mass enclosed is the same as what it would be in FRW, namely
\begin{equation}
M=\frac{H_{0}^{2}\Omega_{M}R^{3}}{2G}.
\label{eq:3}
\end{equation}
We denote by $f(t)$ the fraction of this mass in the thin shell, so
that the mass in the interior is $[1-f(t)] M$.  The fractional density
perturbation in comoving coordinates $\delta_{\rm m}({\bf x},t)=
\delta\rho({\bf x},t)/\rho$ is therefore
\begin{equation}
\delta_{\rm m}({\bf x},t)=- f(t) \Theta(R-r)+
\frac{1}{3} f(t) R
\delta(r-R),\label{eq:11}\end{equation}
where $\Theta(x)$ is the function defined by $\Theta(x)=1$ for $x>0$ and $\Theta(x)=0$
for $x<0$.

The corresponding potential perturbation $\phi$, in a Newtonian gauge in which the metric has the form
\begin{equation}
ds^{2}=-\left(1+2\phi\right)dt^{2}+a_{\rm
  ex}^{2}(t)\left(1-2\phi\right)d{\bf x}^{2},\label{eq:9}
\end{equation}
is given by solving the Poisson equation
$\nabla^2\phi=3H_0^{2} \Omega_M \delta_{\rm m}({\bf x},t)/(2 a_{\rm ex})$.
This gives
\begin{equation}
\phi({\bf x},t)=\frac{H_{0}^{2}\Omega_{M} f(t)}{4a_{\rm ex}(t)}
\left(R^{2}-r^{2}\right)\Theta(R-r).\label{eq:12}\end{equation}
The corresponding radial acceleration is
$$
a_r = - \frac{H_{0}^{2}\Omega_{M} f(t)}{2a_{\rm ex}(t)^2}
r \Theta(R-r).
$$
For each void, the potential will take the form (\ref{eq:12}) in a
spherical polar
coordinate system centered on that void, and the total potential is
given by summing over the voids.  The potential vanishes in between the voids.

We next discuss how to compute the time evolution of the fraction $f(t)$ of the void mass in the thin shell.
We will work to leading, Newtonian order in $(H_0 R)^2$, and we will also
neglect the surface pressure that would arise in a relativistic calculation.  The
uniform interior behaves like a positive energy FRW cosmology. It
has negative curvature, $k<0$, and a scale factor $a_{\rm in}(t)$
that obeys the equation
\begin{equation}
\left(\frac{\dot{a}_{\rm in}}{a_{\rm
      in}}\right)^{2}=H_{0}^{2}\left(\frac{\Omega_{M}}{a_{\rm
      in}^{3}}+1-\Omega_{M}-\frac{k}{a_{\rm in}^{2}H_{0}^{2}}\right),
\label{eq:4}
\end{equation}
since the cosmological constant is the same inside and outside the
void but the matter density is not. We define the positive constant
$a_{0}=-\Omega_{M}H_{0}^{2}/k$, the inverse of which is proportional
to the density contrast at early times. The solution to Eq.\ (\ref{eq:4}) is
\begin{equation}
\frac{3H_{0}t\sqrt{1-\Omega_{M}}}{2}=\int_{0}^{\left(\frac{a_{\rm
        in}}{a_{\Lambda}}\right)^{\frac{3}{2}}}
\frac{dx}{\sqrt{1+x^{2}+x^{\frac{2}{3}}\frac{a_\Lambda}{a_{0}}}}.
\label{eq:5}
\end{equation}
This solution assumes that $a_{\rm in}=a_{\rm ex} = 0$ at $t=0$, so that the
interior and the exterior regions started expanding at the same time.
Otherwise the deviations from FRW are large at early times. Eliminating
$t$ between Eqs.\ (\ref{eq:2}) and (\ref{eq:5}) gives the relationship
between $a_{\rm in}$
and $a_{\rm ex}$, which is
\begin{equation}
\sinh^{-1}\left(\frac{a_{\rm
      ex}^{3/2}}{a_{\Lambda}^{3/2}}\right)=\int_{0}^{\left(\frac{a_{\rm
        in}}{a_{\Lambda}}\right)^{\frac{3}{2}}}
\frac{dx}{\sqrt{1+x^{2}+x^{\frac{2}{3}}\frac{a_\Lambda}{a_{0}}}}.
\label{eq:6}
\end{equation}
Note that the above equations imply that $a_{\rm in}>a_{\rm ex}$, as $k<0$.
The density of the interior is equal to the mean FRW density times
$(a_{\rm ex}/a_{\rm in})^{3}<1$, and so the fraction of mass in the shell is
\begin{equation}
f(t) = 1-\left(\frac{a_{\rm ex}}{a_{\rm in}}\right)^{3}.
\label{eq:8a}
\end{equation}
We numerically solve Eq.\ (\ref{eq:6}) to obtain $a_{\rm ex}/a_{\rm in}$
as a function of $a_{\rm ex}/a_{\Lambda}$, $\Omega_{M}$ and $a_{0}$.
In the remainder of the paper, we will parameterize our void models
in terms of the value today $f_0 = f(t_0)$ of the mass fraction $f(t)$ in the
shell.  We will usually pick $f_0 = 0.9$.  The parameter $a_0$ can be
computed from $f_0$ and $\Omega_M$.

\subsection{Algorithm for randomization of void placement}
\label{sec:placement}

We now discuss how we choose the number and locations of voids in
our model.  In some previous studies \cite{Marra1,Marra2,Marra3}, the
centers of
all the voids encountered by a given ray were chosen to be collinear,
so that the ray passed through the centers of all the voids.
In these studies the lensing demagnification was large enough to
successfully mimic the effects of dark energy.  However, as discussed
in VFW08, the large demagnification was an artifact of
the non-randomness of the void locations, which is not in accord with
observations of the distribution of voids
\cite{voids1,voids2,voids3,voids4}.  In this paper, we use a more
realistic void distribution,
which we compute according to the following procedure:

\begin{enumerate}
\item Fix the comoving void size $R$.
\item Fix the redshift of the source $z_s$.
\item Place voids all along the ray from the source to the observer,
lined up so that they are just touching. The source and the observer
are placed in FRW regions. The distance from the source to the shell
of the adjacent void is chosen to be a fixed small parameter, and the
distance between the observer and the shell of the adjacent void then
depends on the number of voids that can fit between the source and
observer.
\item Randomize impact parameters by shifting each void in a random
  direction perpendicular to the direction of the light ray, so that
  the square $b^2$ of the impact parameter is uniformly distributed
  between $0$ and $R^2$.
\end{enumerate}

Note that with this algorithm, each ray spends some time in FRW
regions between each pair of voids.
An alternative procedure would that used by Holz \& Wald \cite{HW}, in
which after exiting a void, a ray immediately enters another void
without traversing an FRW region.  In this model the effective packing
fraction of voids would be a factor $\sim 2$ or so higher than in our
model, and the rms magnifications and demagnification would be
correspondingly enhanced.

\subsection{Method of computing magnification along a ray}
\label{sec:method}

We now turn to a description of the method we use to compute the magnification
for a ray propagating through a Universe filled with randomly placed
voids, as described in the last subsection. Our method is essentially
a modification of the method introduced by Holz \& Wald \cite{HW},
and goes beyond weak-lensing theory. In this section we describe the
computational procedure; a derivation is given in Appendix \ref{appB}.

Starting from the perturbed FRW metric (\ref{eq:9}), we consider an observer
at $t=t_{0}$ (today) and $\mathbf{x}=0$, or equivalently at $\eta=\eta_{0}$,
where $\eta=\int dt/a_{\rm ex}\left(t\right)$ is conformal time. We consider
a source at $\mathbf{x}=\mathbf{x}_{s}=x_{s}\mathbf{n}$, where $\mathbf{n}$
is a unit vector. The geodesic joining the source and observer in
the background FRW geometry is\begin{equation}
x^{\alpha}\left(x\right)=\left(\eta_{0}-x,\:\mathbf{n}x\right),\label{eq:13}\end{equation}
for $0\leq x\leq x_{s}$, where $x$ is the comoving distance
(or affine parameter with respect to the flat metric  $d\bar{s}^{2}=
a_{\rm ex}\left(t\right)^{-2}ds^{2}=-d\eta^{2}+d\mathbf{x}^{2}$).
Following Holz \& Wald \cite{HW}, we solve the geodesic deviation
equation relative to this unperturbed ray in order to find the net
magnification and shear.  We do not include deflection of the central
ray since the resulting corrections are relatively small; see Appendix
\ref{appB} and Ref.\ \cite{HW}.

We denote by $\vec{k}=d/dx=-\partial_{\eta}+n^{i}\partial_{i}$ the
past directed tangent vector to the ray.
We also introduce a pair of spatial basis vectors ${\vec e}_A$, $A =
1,2$, so that ${\vec e}_A$ and ${\bf n}$ are orthonormal with respect
to $d {\bar s}^2$.
We define the projected
Riemann tensor\begin{equation}
{\cal R}_{AB}=\bar{R}_{\alpha\gamma\beta\delta}k^{\gamma}k^{\delta}e_{A}^{\alpha}e_{B}^{\beta},\label{eq:14}\end{equation}
for $A,B=1,2$ where $\bar{R}_{\alpha\gamma\beta\delta}$ is
the Riemann tensor of the perturbed FRW metric without the $a_{\rm
  ex}\left(t\right)^{2}$ factor:
\begin{equation}
ds^{2}=-\left(1+2\phi\right)d\eta^{2}+\left(1-2\phi\right)d\mathbf{x}^{2}.
\label{eq:15}
\end{equation}

Next we consider the differential equation along the ray
\begin{equation}
\frac{d^{2}}{dx^{2}}\mathcal{A}_{\:\:\: B}^{A}\left(x\right)=
-{\cal R}_{\:\:\: C}^{A}\left(x\right)\mathcal{A}_{\:\:\:
  B}^{C}\left(x\right),
\label{eq:16}
\end{equation}
where ${\cal R}_{\:\:\: C}^{A}\left(x\right)$ means the projected Riemann
tensor evaluated at $x^{\alpha}=x^{\alpha}\left(x\right)$, and capital
Roman indices are raised and lowered with $\delta_{AB}$. We solve
the differential equation (\ref{eq:16}) subject to the initial conditions at
the observer
\begin{equation}
\mathcal{A}_{\:\:\:
  B}^{A}\left(0\right)=0,\:\:\:\:\:\frac{d\mathcal{A}_{\:\:\:
    B}^{A}}{dx}\left(0\right)=\delta_{\:\:\: B}^{A}.
\label{eq:17}
\end{equation}
Finally the magnification along the ray, relative to the background
FRW metric, is\footnote{In our Monte Carlo simulations we discard all
  cases where the determinant is negative, and so the absolute value
  sign in Eq.\ (\ref{eq:18}) can be dropped.  As explained in Ref.\
  \cite{HW}, this prescription yields the distribution of
  magnifications of primary images; it is not possible using the
  geodesic deviation equation method to compute the distribution of
  total luminosity of all the images of a source.
}
\begin{equation}
\mu=\frac{x_s^{2}}{|\det \boldsymbol{{\cal A}}(x_s)|},\label{eq:18}
\end{equation}
where the right hand side is evaluated at the location $x=x_{s}$
of the source.  Note that this quantity, the ratio between the perturbed and unperturbed angular diameter distances, is a conformal invariant, as we show explicitly in Appendix \ref{appB}.

The matrix $\boldsymbol{{\cal A}}(x_s)/x_s$ can be expressed as a
product of an orthogonal matrix and a symmetric matrix with two real
eigenvalues $1 -\kappa \pm \gamma$, where $\kappa$
is called the lensing convergence and $\gamma$ the shear. The
magnification is therefore
\be
\mu = | (1-\kappa)^2 - \gamma^2|^{-1}.
\label{mu00}
\ee

This computational procedure is essentially the same as that used
by Holz \& Wald \cite{HW}, except that Holz \& Wald work in the
physical spacetime rather than the conformally transformed spacetime,
and at
the end of the computation they compute the ratio between the quantity
$x_{s}^{2}/(\det \boldsymbol{{\cal A}})$ evaluated
in the perturbed spacetime and in the background spacetime. In our
approach we do not need to compute a ratio, and furthermore the source
term in the differential equation (\ref{eq:16}) vanishes in FRW regions between
the voids, which simplifies the computation. See Appendix \ref{appB}
for more details on the relation between the two approaches.

We now turn to a discussion of the method we use to compute approximate
solutions to the differential equation (\ref{eq:16}).
Consider a small segment of ray, from $x=x_{1}$ to $x=x_{2}$
say. Since the differential
equation is linear, we have
\bea
\left[\begin{array}{c}
\mathcal{A}_{\:\:\: B}^{A}(x_2)\\
\dot{\mathcal{A}}_{\:\:\: B}^{A}(x_2)\end{array}\right]
&=&
\left[\begin{array}{cc}
J_{\:\:\: C}^{A}\left(x_{2},\; x_{1}\right) & K_{\:\:\: C}^{A}\left(x_{2},\; x_{1}\right)\\
L_{\:\:\: C}^{A}\left(x_{2},\; x_{1}\right) & M_{\:\:\:
  C}^{A}\left(x_{2},\; x_{1}\right)\end{array}\right]
\nonumber \\
\mbox{} && \times
\left[\begin{array}{c}
\mathcal{A}_{\:\:\: B}^{C}(x_1)\\
\dot{\mathcal{A}}_{\:\:\: B}^{C}(x_1)\end{array}\right].
\label{eq:26}
\eea
for some $2\times2$ matrices $J,\: K,\: L,\: M$ which together form
a $4\times4$ matrix.  To linear order in ${\cal R}_{AB}$ we
have\footnote{Holz \& Wald \protect{\cite{HW}} drop all of the
  integrals over the projected Riemann tensor in Eqs.\
  (\protect{\ref{JKLM}}) except the one in the formula for $L^A_{\
    B}$.  This is valid to leading order in $(H_0 R)^2$.
We keep the extra terms in Eqs.\ (\protect{\ref{JKLM}})
even though our formalism neglects other
effects that also give fractional corrections of order $(H_0 R)^2$.
The extra terms change $\sigma_m$ by a few percent.}
\bes
\label{JKLM}
\bea
\label{eq:21}
J_{\:\:\: C}^{A}&=&\delta_{\:\:\:
  C}^{A}-\int_{x_{1}}^{x_{2}}dx\left(x_{2}-x\right){\cal R}_{\:\:\:
  C}^{A}\left(x\right), \\
\label{eq:22}
K_{\:\:\: C}^{A}&=&\left(x_{2}-x_{1}\right)\delta_{\:\:\:
  C}^{A}\nn \\
&&-\int_{x_{1}}^{x_{2}}dx\int_{x_{1}}^{x}d\bar{x}\left(\bar{x}-x_{1}\right){\cal
  R}_{\:\:\: C}^{A}\left(\bar{x}\right),\ \  \\
\label{eq:23}
L_{\:\:\: C}^{A}&=&-\int_{x_{1}}^{x_{2}}dx{\cal R}_{\:\:\:
  C}^{A}\left(x\right), \\
\label{eq:24}
M_{\:\:\: C}^{A}&=&\delta_{\:\:\:
  C}^{A}-\int_{x_{1}}^{x_{2}}dx\left(x-x_{1}\right){\cal R}_{\:\:\:
  C}^{A}\left(x\right).
\eea
\ees
We evaluate these matrices for a transition through a single void,
using the potential (\ref{eq:12}), the metric (\ref{eq:15}) and the
definition
(\ref{eq:14}) of ${\cal R}_{AB}$.
We neglect the time evolution of the potential during passage through
the void; the corresponding corrections are suppressed by $(H_0 R)^2$.  This gives
\bes
\label{JKLM1}
\bea
\label{eq:21a}
J_{\:\:\: C}^{A}&=&
\delta_{\:\:\: C}^{A}+c^2 {\cal
  P}(z)\left(\begin{array}{cc}
1 & 4\\
4 & 1\end{array}\right),\\
\label{eq:22a}
K_{\:\:\: C}^{A}&=&
(x_{2}-x_{1})\delta_{\:\:\: C}^{A}+\frac{2}{3}c^3 {\cal P}(z) \left(\begin{array}{cc}
1 & 2\\
2 & 1\end{array}\right),\ \
\\
\label{eq:23a}
L_{\:\:\: C}^{A}&=&
2 c {\cal
  P}(z)\left(\begin{array}{cc}
1 - \frac{R^2}{3 c^2}& 4\\
4 & 1 - \frac{R^2}{3 c^2}\end{array}\right),\ \ \  \ \ \ \\
\label{eq:24a}
M_{\:\:\: C}^{A}&=&
\delta_{\:\:\: C}^{A}
+2c^2 {\cal P}(z)\left(\begin{array}{cc}
1 + \frac{R^2}{3 c^2}
 & 2\\
2 & 1+\frac{R^2}{3 c^2}\end{array}\right).\ \ \ \ \ \
\eea
\ees
Here $b$ is the impact parameter, $c = \sqrt{R^2 - b^2}$,
\begin{equation}
{\cal P}(z)=\frac{3}{2}H_{0}^{2}\Omega_{M}\frac{x\left(x_{s}-x\right)}{x_{s}
} \frac{f(z)}{a_{\rm ex}(z)},
\label{eq:33-1}
\end{equation}
and $f(z)$ is defined by Eq.\ (\ref{eq:8a}).
In these equations $x$ and $z$ are evaluated at the center of the
void.

Our computational procedure can now be summarized as follows:

\begin{enumerate}

\item Pick some source redshift $z_s$, void radius $R$, and fraction
  of void mass on the shell today $f_0$.

\item Choose void locations according
to the prescription described in Sec.\ref{sec:placement}.

\item For each void, compute the $4\times4$ matrix that is
formed by the matrices ${\bf J}, {\bf K}, {\bf L}$ and ${\bf M}$
from Eqs.\ (\ref{JKLM1}).

\item Perform a similarity
transformation ${\bf J} \to {\bf U}^{-1} \cdot {\bf J} \cdot {\bf U}$
on each of the matrices ${\bf J}, {\bf K}, {\bf L}, {\bf M}$ for some
randomly chosen $SO(2)$ matrix ${\bf U}$, to randomize the
direction of the vectorial impact parameter.

\item Multiply together all the $4\times4$ matrices, and multiply by
  the initial conditions (\ref{eq:17}), to evaluate
  $\mathcal{A}_{\:\:\: B}^{A}\left(x_{s}\right)$.

\item
Compute the magnification $\mu$ relative to FRW from Eq.\
(\ref{eq:18}), and then distance modulus shift $\Delta m$ from
\bes
\bea
\label{eq:25a}
\Delta m&=&-\frac{5}{2}\log_{10}\left(\mu\right)\\
&=& {5 \over 2 \ln 10} \ln | (1-\kappa)^2 - \gamma^2 |.
\label{eq:25}
\eea
\ees

\item Repeat steps 2 to 6 a large number of times to generate the
  distribution $p(\Delta m;z_s)$ of distance modulus shifts $\Delta m$
for sources at redshift $z_s$, for a randomly chosen direction from
the observer.

\item Finally, we correct this distribution to obtain the
  observationally relevant quantity, the probability distribution of
magnitude shifts for a source chosen randomly on a sphere at a
distance corresponding to redshift $z_s$.  The corrected distribution
is \cite{HW}
\bea
{\cal P}(\Delta m;z_s) &=& {\cal N} p(\Delta m;z_s)/\mu
\nn \\
&=& {\cal N} p(\Delta m;
z_s) 10^{2 \Delta m/5},
\label{OR}
\eea
where ${\cal N}$ is a normalization constant.

\end{enumerate}

\subsection{Relation to weak lensing theory}

In weak lensing theory the matrix $\boldsymbol{\cal A}(x_s)/x_s$ that
describes the deflections of the rays is presumed to be always very
close to the unit matrix, so the total integrated effect of the inhomogeneities
on a given ray can be treated linearly.  The solution to
Eq.\ (\ref{eq:16}) in this approximation is given by Eq.\ (\ref{eq:22})
with $x_1 =0$, $x_2=x_s$,
\be
\frac{{\cal A}^A_{\:\:\: B}(x_s)}{x_s} = \delta^A_{\:\:\: B} -
\int_0^{x_s} dx \, \frac{x(x_s-x)}{x_s} {\cal R}^A_{\:\:\: B}(x).
\ee
Taking the determinant and linearizing again, the contribution from
shear vanishes and the magnification is $\mu = 1 + 2 \kappa$ where
the lensing convergence $\kappa$ is given by the standard formula
\begin{equation}
\kappa=\frac{3}{2}H_{0}^{2}\Omega_{M}\int_{0}^{x_{s}}dx\frac{x\left(x_{s}-x\right)}{x_{s}a_{\rm
    ex}\left(z\right)}\delta_{\rm m}(x).
\label{eq:42-1}
\end{equation}
Here $\delta_{\rm m}(x)$ is the fractional over
density, $x$ is the comoving distance, $x_{s}$ is comoving distance
to the source, and $a_{\rm ex}\left(z\right)$ is the scale factor.
Evaluating this for our void model gives
\be
\kappa = \sum_i \kappa_i,
\label{WLR0}
\ee
where the sum is over the voids and
\be
\kappa_i = - 3 H_0^2 \Omega_M \frac{x_i (x_s - x_i)}{x_s a_{\rm
    ex}(z_i)} f(z_i) c_i \left[ 1 - \frac{R^2}{3 c_i^2} \right]
\label{WLR}
\ee
is the lensing convergence from the $i$th void.
Here $z_i$ and $x_i$ are the redshift and
comoving distance to the center of the $i$th void, $c_i = \sqrt{R^2 -
  b_i^2}$ and $b_i$ is the $i$th impact parameter.
Our model goes beyond the weak lensing result (\ref{WLR0}) as it
includes lens-lens couplings and shear.

\section{Approximate Analytical Computation of magnification dispersion}
\label{sec:analyticalmodel}

\subsection{Overview}
\label{sec:modeloverview}

In the previous section, we described a Monte Carlo procedure for
computing the probability distribution ${\cal P}(\Delta
m; z_s)$ of magnitude shifts $\Delta m$ for sources at redshift $z_s$,
for our Swiss cheese model of voids.
We will be particularly interested in the mean
\be
\left< \Delta m \right> = \int d\Delta m \, \Delta m {\cal P}(\Delta m; z_s)
\ee
and variance
\be
\sigma_m^2 = \int d\Delta m \, \left( \Delta m
- \left< \Delta m \right> \right)^2 {\cal P}(\Delta m; z_s)
\ee
of this distribution.
In subsequent sections of the
paper we will describe the results of our Monte Carlo simulations and
their implications.  In this section, however, we will take a detour
and describe a simple, approximate, analytic computation of the
variance.  The approximation consists of
using the weak lensing approximation to compute the total lensing
convergence $\kappa$ (accurate to a few percent, see Sec.\
\ref{sec:nonlin}),
and then using an approximate cutoff procedure
to incorporate the effect of the nonlinear relation (\ref{eq:25})
between $\kappa$ and the magnitude shift $\Delta m$.
We will see
in Sec.\ \ref{sec:results}
below that this analytic approximation agrees with our
Monte Carlo simulations to within $\sim 30\%$.

Neglecting shear, the relation (\ref{eq:25}) reduces to
\be
\Delta m = {5 \over \ln 10} \ln | 1 - \kappa |
\label{eq:25b}
\ee
where $\kappa$ is given by Eqs.\ (\ref{WLR0}) and (\ref{WLR}).  We will see
shortly that the the variance of $\kappa$ diverges. This divergence is
an artifact of our use of a distributional density profile for each
void, with a $\delta$-function on the void's surface, and can be
removed by endowing each shell with some small finite thickness
$\Delta r$ (see Sec.\ \ref{sec:finite} below).  The variance of $\Delta m$, on
the other hand, is finite,
because of the nonlinear relation (\ref{eq:25b}).  We shall proceed by
using the linearized version
\be
\Delta m = - {5 \over \ln 10} \left[ \kappa + O(\kappa^2) \right]
\label{eq:25c}
\ee
of Eq.\ (\ref{eq:25b}), and by simply
cutting off the divergent integrals that arise, at $\kappa\sim 1$, the
regime where the nonlinearity of the relation (\ref{eq:25b}) becomes
important.

\subsection{Variance of magnitude shifts}
\label{sec:variance}

From Eq.\ (\ref{eq:25c}) we find for the mean and variance of the
magnitude shift
\ba
\langle\Delta m\rangle&=& - {5\over\ln 10}\left[\langle\kappa\rangle
+O(\kappa^2)\right],
\nonumber\\
\sigma_m^2&=&\left({5\over\ln 10}\right)^2\left[ \langle\kappa^2\rangle-
\langle\kappa\rangle^2 + O(\kappa^3) \right].
\label{meanandvariance}
\ea
The averages are over the set of impact parameters $\{b_i:i\in
[1,j(x_s)]\}$ in Eq.\ (\ref{WLR}),
where $j(x_s)$ is the number of voids out to the source at $x_s$.
In computing the averages, it will prove convenient to define
\be
q_i=1-b_i^2/R^2,
\label{qidef}
\ee
so that each $q_i$ is distributed uniformly between zero and one,
since impact parameters arbitrarily close to the void boundary are permitted.
In fact, a shortcoming of our model is the vanishing thickness of the void
wall.  We therefore introduce lower cutoffs $C_i$ for each
void\footnote{These cutoffs will be used only for construction of
  our analytical model in this section; they are not used in Monte
  Carlo simulations in the remainder of the paper.}, that is, we
restrict $q_i$ to lie in the range $C_i \le q_i
\le 1$.  We will discuss below the origin and appropriate values of
these cutoffs.

With this assumption we obtain for the mean of the  lensing
convergence (\ref{WLR}) of the $i$th void
\bea
\langle\kappa_i\rangle &=&-{H_0^2\omm x_sR w_i}
\int_{C_i}^1 dq_i \left[
3\sqrt{q_i}-{1\over\sqrt{q_i}} \right]\nn \\
&=&-{2H_0^2\omm x_sR w_i\sqrt{C_i}(1-C_i)}
\label{kappamean}
\eea
where
\be
w_i={x_i(x_s-x_i)f(z_i)(1+z_i)\over x_s^2}~.
\label{widef}
\ee
The mean lensing convergence (\ref{kappamean}) is always negative,
since $C_i<1$; introducing
the cutoff leads to a bias toward de-magnification. This is a shortcoming
of the model, since for any mass-compensated perturbation $\langle\kappa_i
\rangle=0$ \footnote{To restore this feature we could either scale the contribution
from the underdense core
downward by a factor of $S_i=1+\sqrt{C_i}+C_i$ or scale the contribution
from the overdense shell upward by the same factor $S_i$.}. Below, we shall
ignore small corrections that are powers of $C_i$, and will take
$\langle\kappa_i\rangle=0$ for all $i$.

By contrast the second moment $\langle\kappa_i^2\rangle$ diverges
logarithmically in the limit $C_i \to 0$:
\bes
\bea
\label{kappameansquare1}
\langle\kappa_i^2\rangle &=& \left({H_0^2\omm x_sR w_i}\right)^2
\int_{C_i}^1 dq_i
\left(3\sqrt{q_i}-{1\over\sqrt{q_i}}\right)^2  \ \ \ \
 \\
&=&\left({H_0^2\omm x_sR w_i}\right)^2\left[-\ln C_i-{3\over 2}+
  O(C_i)
\right].\ \ \
\label{kappameansquare}
\eea
\ees
This divergence is caused by rays that just graze the $\delta$
function shell of the void.

Because $\kappa$ is a sum of $\kappa_i$, its mean is the sum of the individual
means, but
\be
\langle\kappa^2\rangle=\sum_i\langle\kappa_i^2\rangle-
\sum_{i\neq j}\langle\kappa_i\rangle\langle\kappa_j\rangle
\ee
and therefore
\be
\sigma_\kappa^2=\langle\kappa^2\rangle-\langle\kappa\rangle^2=
\sum_i(\langle\kappa_i^2\rangle-\langle\kappa_i\rangle^2)~.
\ee
Combining this with Eqs.\ (\ref{eq:25c}) and (\ref{kappameansquare})
and dropping terms linear in $C_i$ gives
for the variance in magnitude shift
\be
\sigma_m^2 = \sigma_0^2 \sum_i
w_i^2\left(-\ln C_i-{3\over 2}\right),
\label{analytic0}
\ee
where we have defined
\be
\sigma_0 = {5H_0^2\omm x_sR\over\ln 10}.
\ee
We choose the cutoffs $C_i$ to correspond to $\kappa_i \sim 1$, as
discussed above; from Eqs.\ (\ref{WLR}) and (\ref{widef}) this gives
\be
C_i = (H_0^2\omm Rx_sw_i)^2.
\label{cutoff1}
\ee
The approximate analytic result given by Eqs.\ (\ref{widef}) and
(\ref{analytic0}) -- (\ref{cutoff1}) is plotted in Fig.\ \ref{fig:nonlinear} in
Sec.\ \ref{sec:nonlin} below.  It agrees with our Monte Carlo simulations to
within $\sim 30\%$, which is reasonable given the crudeness of our
analytic cutoff procedure.

\subsection{Finite sampling effects}

In addition to computing the width $\sigma_m^2$ of the distribution of
magnitude shifts $\Delta m$, we now compute a different quantity
$\sigma_{m,{\rm med}}^2(N)$ which
is, roughly speaking, the estimate of the width that one would obtain
with $N$ samples $\Delta m_\alpha$, $1\le \alpha \ne N$, drawn from
the distribution.  More precisely, this quantity is defined as
follows.  From the $N$ samples we construct the estimator
\be
{\hat \sigma}_m^2 \equiv {1 \over N-1} \sum_{\alpha=1}^N \Delta
m_\alpha^2 - {1 \over N(N-1)} \left( \sum_{\alpha=1}^N \Delta m_\alpha
\right)^2.
\label{estimator}
\ee
This quantity is itself a random variable with expected value $\left<
  {\hat \sigma}_m^2 \right> = \sigma_m^2$.  However for finite $N$ the
median value of the distribution of ${\hat \sigma}_m^2$ can be significantly different from
$\sigma_m^2$.
We denote this median value by $\sigma_{m,{\rm med}}^2(N)$.
In the limit $N \to \infty$ we have $\sigma_{m,{\rm med}}(N) \to
\sigma_m$.
We note that realistic supernovae surveys will have
no more than $\sim 10^4$ supernovae.

To estimate this median value, we use the fact that
for each void $i$, finite sampling imposes a minimum value on $q_i$ of
$q_i \sim 1/N$ on average, which acts like a statistical cutoff in the integral
(\ref{kappameansquare}).
This is in addition to the physical cutoff (\ref{cutoff1}) discussed above,
which we will denote by $q_{i,c}$ from now on.
For $N$ samples $q_{i,\alpha}$, $1 \le \alpha \le N$,
the probability that all $N$ samples are larger than a value $C_i$
which is larger than $q_{i,c}$ is
\be
P_0(<C_i)=\left({1-C_i\over 1-\qci}\right)^N.
\ee
Differentiating once we find that the probability distribution of $C_i$
is
\be
P(C_i)=\left\vert{dP_0(<C_i)\over dC_i}\right\vert=
{N(1-C_i)^{N-1}\over (1-\qci)^N}~.
\ee
For very large values of $N$ and small $\qci$ an adequate approximation is
\be
P(C_i)\approx N\exp[-N(C_i-\qci)],
\label{PCi}
\ee
which is properly normalized for $C_i\geq\qci$
if we extend the range of $C_i$ to infinity, thereby incurring an
error $\sim \exp(-N)$.

We now average the expression (\ref{analytic0}) for the width $\sigma_m^2$, using the
distribution (\ref{PCi}) to average over the cutoffs $C_i$.  The
result is
\bea
\label{sigavgd}
\sigma_{m,{\rm med}}^2 &\approx&\sigma_0^2
\sum_i
w_i^2
 \\ &&\times
\left[\ln N-{3\over 2}
 -\int_0^\infty dxe^{-x}\ln(x+N\qci)\right]. \nn
\eea
If we define
\ba
\label{Sdef}
S(f_0,z_s)&=&\sum_iw_i^2,
\nonumber\\
\gamma(N\qci)&=&-\int_0^\infty dxe^{-x} \ln(x+N\qci),
\label{Sgammadef}
\ea
then Eq.\ (\ref{sigavgd}) becomes
\bea
\sigma_{m,{\rm med}}^2 &\approx& \sigma_0^2
\left[S(f_0,z_s)\left(\ln N-{3\over 2}\right)+\sum_i w_i^2\gamma(N\qci)
\right].
\nn \\ &&
\label{sigavgdp}
\eea

The result (\ref{sigavgdp}) was obtained by averaging over the cutoffs
$\{C_i\}$ using the probability distribution (\ref{PCi}), and is an
estimate of the median of the distribution of ${\hat \sigma}_m^2$.
Of course the actual value of ${\hat \sigma}_m^2$ computed from a Monte Carlo
realization of $N$ lines of sight, or
obtained from $N$ observations of magnifications,
may differ from the result (\ref{sigavgdp}).
We would like to also estimate the spread in values of ${\hat \sigma}_m^2$.
From Eq.\ (\ref{analytic0}), and taking the variance with respect to
the distribution of cutoffs $C_i$, we find
\be
\left( { \Delta \sigma_{m,{\rm med}}^2 \over \sigma_{m,{\rm med}}^2}
\right)^2 = { \sum_i w_i^4 \Var(N \qci) \over \left[ \sum_i w_i^2
    ({\overline{ \ln C_i}} + 3/2) \right]^2},
\label{eq:spread}
\ee
where
\bea
\Var(N\qci)&=&{\overline{(\ln C_i)^2}}-\left({\overline{\ln C_i}}\right)^2
\nn \\
&=&\int_0^\infty dxe^{-x}[\ln(x+N\qci)]^2-[\gamma(N\qci)]^2\nn\\&&
\label{varlnCi}
\eea
Here the overbars denote an expectation value with respect to the
probability distribution (\ref{PCi}).
The quantity (\ref{eq:spread}) is a measure in the fractional spread
in our estimate of the median,
and should give a lower bound on the fractional spread in values of
${\hat \sigma}_m^2$.

Two limits of Eqs.\ (\ref{sigavgdp}) and (\ref{eq:spread})
are especially simple.  First, for $N\qci\ll 1$, we have
$\gamma(N\qci)\approx\gamma_E=0.5772\ldots$, the Euler-Mascheroni
constant, and also $\Var(N\qci) \approx 1.645$ and $-{\overline{\ln
    C_i}} \approx \ln N + \gamma_E$.  This gives
\bes
\label{smallqci}
\ba
\sigma_{m,{\rm med}}^2&\approx&\sigma_0^2
S(f_0,z_s)\left(\ln N-{3\over 2}+\gamma_E\right),\ \ \
\label{smallqci1}
\\
{ \Delta \sigma_{m,{\rm med}}^2 \over \sigma_{m,{\rm med}}^2}
 &\approx& \sqrt{ 1.645 \over N_{\rm void} (\ln N + \gamma_E - 3/2) },
\label{smallqci2}
\ea
\ees
where $N_{\rm void} = x_s/(2 R)$ is the number of voids
and we have used the crude approximation $w_i = $ constant in the
second equation.
Second, for $N\qci\gg 1$, we have $\gamma(N\qci)\approx -\ln N\qci$,
$\Var(N\qci) \approx 1/ (N\qci)^2$,
and ${\overline{ \ln C_i}} \approx \ln \qci$,
and so we obtain
\bes
\label{largeqci}
\ba
\sigma_{m,{\rm med}}^2&\approx&\sigma_0^2
\left[-{3 \over 2} S(f_0,z_s)-\sum_iw_i^2\ln \qci\right]\ \ \ \
\label{largeqci1}
\\
{ \Delta \sigma_{m,{\rm med}}^2 \over \sigma_{m,{\rm med}}^2} &\propto&
{1 \over N}.
\label{largeqci2}
\ea
\ees
The second case (\ref{largeqci1}) coincides with the $N$-independent
width (\ref{analytic0}) -- (\ref{cutoff1}) computed earlier.
We see that the results are dictated by a competition between statistical
and physical cutoffs via the dimensionless parameter $N\qci$.

\begin{figure}
\bec
\ifx\dofigures\undefined
\else
  \includegraphics[scale=0.55]{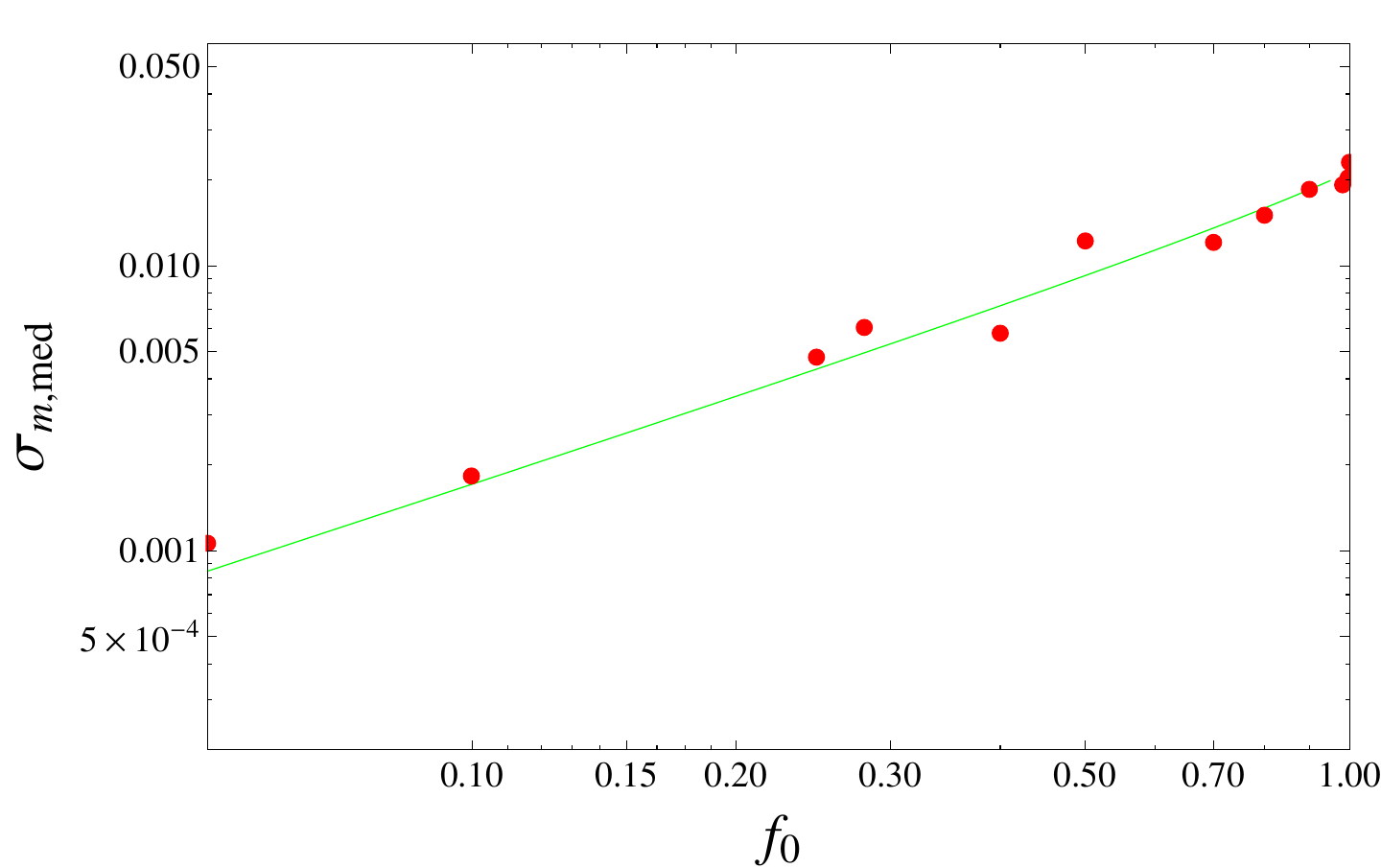}
\fi
\caption{The green line is our analytic model (\ref{smallqci1}) of the
  median width of the distribution
  of magnitude shifts $\Delta m$, for $N = 10^4$ samples, source
  redshift $z_s = 1.0$ and void radius $R = 35$ Mpc, as a function of
  the fraction of mass $f_0$ on the void shells today.  The data
  points are from our Monte Carlo simulations with the same parameter
  values, described in Sec.\ \ref{sec:results} below.}
\label{fig:Xb}
\eec
\end{figure}

As discussed above, our simulations are effectively cut off at
$\kappa_i\sim 1$; this implies
a physical cutoff
\bea
\qci&\sim& (H_0^2\omm Rx_sw_i)^2 \nn \\
&\approx&
2.2\times 10^{-7} \left({H_0\Omega_Mx_s \over 0.23}\right)^2
\left({h_{0.7}R\over 35\Mpc}\right)^2(4w_i)^2.\ \ \ \ \ \
\label{estimate1}
\eea
Here we have scaled the factor $\Omega_M H_0 x_s$ to its value at
$\Omega_M = 0.3$, $z_s = 1.0$, the quantity $h_{0.7}$ is given by
$H_0=70h_{0.7}\kmpspmpc$, and we note that $4w_i\leq f(z_i)(1+z_i)$.
From the estimate (\ref{estimate1})
we expect the $N\qci\ll 1$
limit to apply for $N\lesssim 10^6$.  In this case the cutoff is purely
statistical and the physical cutoff is unimportant.  The prediction (\ref{smallqci1})
for $\sigma_{m,{\rm med}}$ for $N = 10^4$ and $z_s = 1$ is shown in
Fig.\ \ref{fig:Xb}, together with results from our Monte Carlo
simulations, which are described in Sec.\ \ref{sec:results} below.
The plot shows good agreement between the model and the simulations.

For this case, a lower bound on the fractional
spread in the values of ${\hat \sigma}_m^2$ around its median value is
given by Eq.\ (\ref{smallqci2}).
That is, in any given simulation or
observational
survey with $N$ light sources, the scatter of values about the
expected will be at least this large.
For example, with $N=10^4$, $z_s = 1$ and $R = 35$ Mpc, the implied
spread is $\agt 6\%$.  In this regime where the cutoff is primarily statistical,
the range of likely values of ${\hat \sigma}_m$ is
substantial, and only
decreases logarithmically with increasing $N$.

When $N\gtrsim 10^6$, we move into the $N\qci\gg 1$ regime where Eqs.
(\ref{largeqci}) apply.
The results in this regime were discussed in Sec.\ \ref{sec:variance} above, and
are plotted in Fig.\ \ref{fig:nonlinear} in
Sec.\ \ref{sec:nonlin} below.
Equation (\ref{largeqci2}) indicates that the spread
scales as $1/N$ in this regime.  However this estimate is
only a lower bound for the spread in values of ${\hat \sigma}_m^2$, as
discussed above.  In fact,
from Eq.\ (\ref{estimator}) the standard deviation
of ${\hat \sigma}_m^2$ can be computed in terms of $N$ and of the second and fourth
  moments of $\Delta m$; it scales like $1/\sqrt{N}$ as $N
  \to \infty$.  In any case, the spread decreases more rapidly as $N$
  increases after the transition to the large $N$ regime.
We will see in Sec.\ \ref{sec:results} below that this prediction agrees
well with our Monte Carlo simulations.

\subsection{Extension of void model to incorporate finite shell thickness}
\label{sec:finite}

In this subsection we consider a modification of our void model,
in which the void wall is given a finite comoving thickness $\Delta r_i$ that
acts as a physical cutoff in the divergent integral
(\ref{kappameansquare1}).  The corresponding value of the cutoff
parameter $\qci$ is $\qci = 2 \Delta r_i/R$, from Eq.\ (\ref{qidef}).
The value of wall thickness that corresponds to the cutoff (\ref{estimate1}) is
thus $\Delta r_i \sim 3 \, {\rm pc} (R / 35 \, {\rm Mpc})^3$,
which is much smaller than
the expected void wall thicknesses $\sim$ Mpc in large scale structure.
Thus, our thin-shell void model is somewhat unrealistic;
the results are modified (albeit only logarithmically) once the wall thickness
exceeds $\sim $ pc scales.  This motivates modifying the model
to incorporate a finite wall thickness.


Consider next how the wall thickness evolves with redshift.  At very
early times, when the perturbation is in the linear regime, it
maintains its shape in comoving coordinates, so the
cutoff scale is some fixed fraction of $R$. Once the perturbation
becomes nonlinear, the shell thickness should freeze out in physical
extent, implying a comoving size $\propto 1/a$. Thus, a suitable model
for the redshift dependence of the cutoff would be
\be
q_c(a)= \epsilon_0 W(a/a_0),
\ee
where $W(x)$ is a function with $W(x)\to 1$ for $x\ll 1$ and $W(x)\to K_0/x$
for $x\gg 1$.  Here $a_0(f_0)$ is the scale factor when the perturbation ceases
to be linear, and $K$ and $\epsilon_0$ are constants that may also depend
on $f_0$. Very roughly, we expect $q_c(a)\sim 0.1$ so $N\qci\gg 1$ as long
as $N\gtrsim 10$, so that Eq.\ (\ref{largeqci}) will apply.

Suppose now that for a
restricted range of source redshifts
it suffices to take the fractional shell wall thickness
$
\epsilon_{\rm s} = \Delta r_i/R
$
in comoving
coordinates to be the same for all shells.  Then
from Eq.\ (\ref{analytic0}) we get\footnote{Eq.\ (\ref{unithick})
differs from Eq.\ (\ref{analytic0}) in that the $-3/2$ has been
replaced by $\ln 2$.  This slightly more accurate version of the
equation is derived as
follows.  Instead of using the cutoff procedure embodied in Eq.\
(\ref{kappameansquare1}), we use a regulated density profile of the
form
$\delta_{\rm m}(r) = - f \Theta(R_1-r) + \alpha
\Theta(R-r) \Theta(r - R_1)$ where $R_1 = R(1 - \epsilon_{\rm s})$ and
$\alpha = f [ (1 - \epsilon_{\rm s})^{-3} -1]^{-1}$.
The variance in the lensing convergence can then be computed from
\bea
\left< \kappa_i^2 \right> &=& 9 H_0^4 \Omega_M^2 {x_i^2 (x_s - x_i)^2
  \over x_s^2 a_{\rm ex}(x_i)^2 R^2}
\int_0^R dr \int_0^R d{\bar r} \delta_{\rm m}(r)
\delta_{\rm m}({\bar r}) \nn \\
&& \times r {\bar r} \ln \left| {r + {\bar r} \over r - {\bar r}} \right|,
\eea
from Eq.\ (\ref{eq:42-1}).}
\be
\sigma_m^2 = \sigma_0^2
S(f_0,z_s)
\left[-\ln\epsilon_{\rm s}+ \ln(2) + O(\epsilon_{\rm s} \ln
  \epsilon_{\rm s}) \right].
\label{unithick}
\ee
Equation (\ref{unithick}) has the same form as Eq.\ (\ref{smallqci1}), but
since $N\epsilon_{\rm s}\gg 1$, the implied $\sigma_m$ is smaller.
For example, evaluating this expression for $f_0 = 0.9$, $z_s = 1.0$ and $R = 35 $ Mpc gives
\be
\sigma_m \approx 0.013 \sqrt{ 1 + 0.23 \ln \left( {1 \, {\rm Mpc} \over \Delta r} \right) }.
\ee
where $\Delta r = \epsilon_{\rm s} R$ is the wall thickness.

The logarithmic divergence of $\sigma_m^2$ will also be regulated by
treating the shell as composed of fragments that represent local density
enhancements such as galaxy clusters and superclusters for purposes of
computing the magnification of passing light beams. We shall examine this
further refinement of our model elsewhere.

\section{Results of Monte Carlo Simulations for Magnification Distributions}
\label{sec:results}

\begin{figure}
\bec
\ifx\dofigures\undefined
\else
  \includegraphics[scale=0.7]{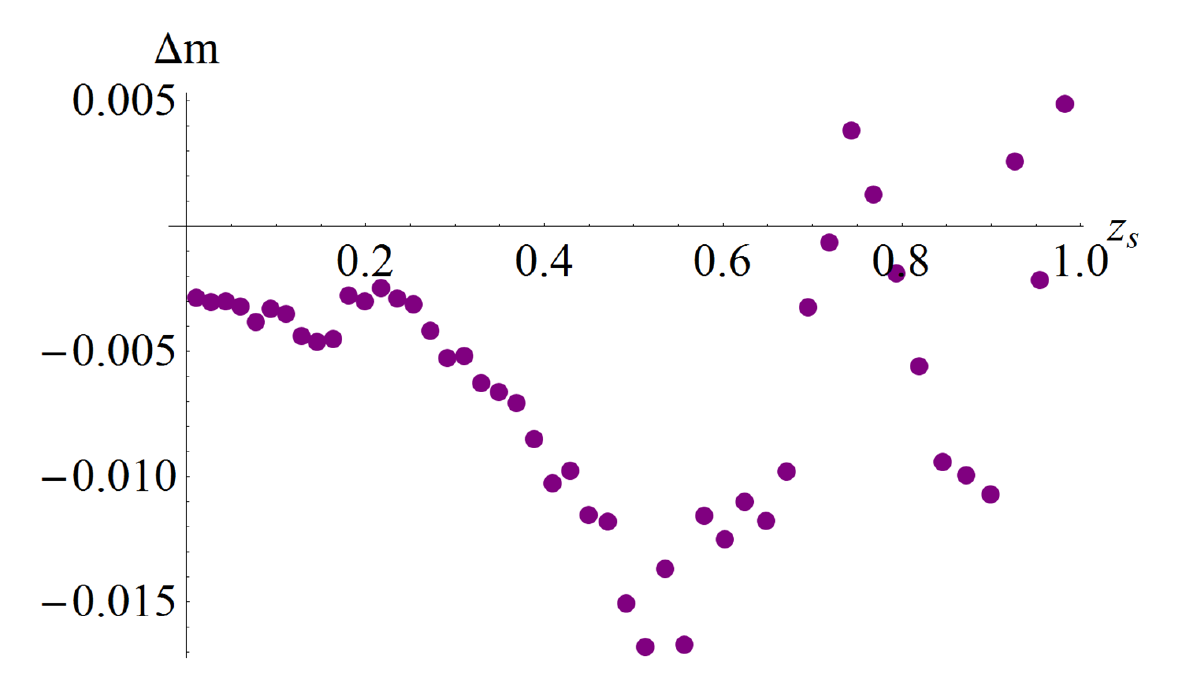}
\fi
\caption{The magnitude shift $\Delta m$ as a function of source
  redshift $z_s$ for a single run, for voids of radius $R = 35$ Mpc, fraction of mass on the shell today $f_0 = 0.9$, in a $\Lambda$CDM cosmology with $\Omega_M = 0.3$.}
\label{fig:OneRun}
\eec
\end{figure}

We now turn to describing the results of our Monte Carlo simulations
based on the algorithm described in Sec.\ \ref{sec:lensingmodel}.
In the remainder of this paper, unless otherwise specified, we will
adopt the fiducial parameter values of matter fraction $\Omega_M =
0.3$, source
redshift $z_s = 1.0$, void size $R = 35 \, {\rm Mpc}$, and fraction of void
mass on shell today $f_0 = 0.9$.
Our choice of void size is motivated by the fact that
observed void sizes
\cite{voids1,voids2,voids3,voids4,voids5,voids6,voids7,voids8,voids9}
range from a typical size of $\sim 10$ Mpc
to an upper limit of $\sim 100$ Mpc.
For this fiducial case, we show in Fig.\ \ref{fig:OneRun}
the distance modulus shift $\Delta m$
as a function of redshift $z_s$ for a single realization of the void
distribution.  The values jump discontinuously after each void,
and illustrate the stochastic nature of the lensing process.

\begin{figure}
\bec
\ifx\dofigures\undefined
\else
\includegraphics[scale=0.7]{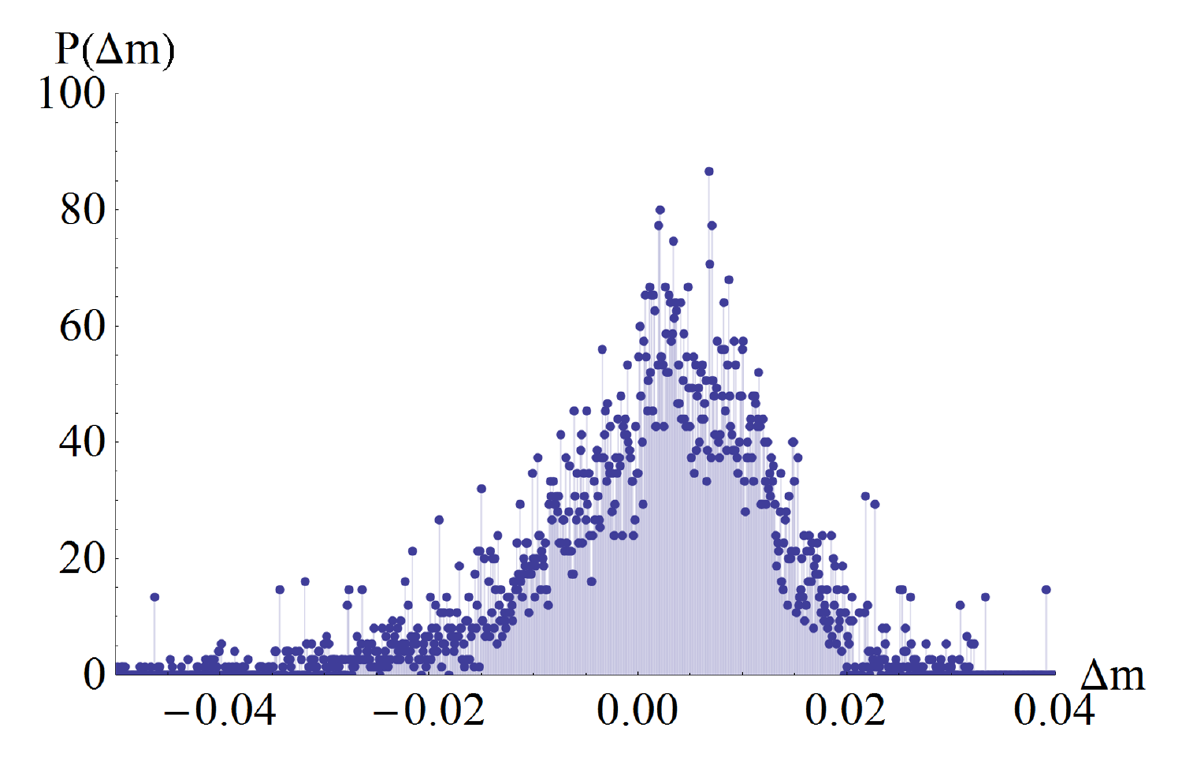}
\fi
\caption{The probability distribution of magnitude shifts $\Delta m$
  for a simulation in a $\Lambda$CDM cosmology with $\Omega_M = 0.3$,
  with sources at redshift $z_s=1$, comoving voids radius
$R = 35$ Mpc, and fraction of void mass on the shell today $f_0 = 0.9$.}
\label{fig:PDF}
\eec
\end{figure}

Next, we repeat this computation some large number $N$ of times in
order to generate the distribution of modulus shifts $\Delta m$.
In the rest of the paper we will focus in particular on the mean
$\left< \Delta m \right>$ and standard deviation $\sigma_m$ of this
distribution, and also on the estimator ${\hat \sigma}_m(N)$
of the standard deviation that one obtains at finite
$N$,
given by Eq.\ (\ref{estimator}), which satisfies ${\hat \sigma}_m(N)
\to \sigma_m$ as $N \to \infty$.

The distribution for the fiducial case for $N=2\times 10^6$
is shown in Fig.\ \ref{fig:PDF}.  For this case the standard deviation is
$\sigma_{m}=0.03135 \pm 0.0003$ and the mean is
$\left\langle \Delta m\right\rangle =0.004 \pm 0.001$
(where the error is estimated based on dividing the data into 200 groups of 10000
runs).
Our result for the standard deviation agrees to within $\sim 30\%$ with
that of a different Swiss cheese void model by Brouzakis, Tetradis and Tzavara
\cite{BTT08}; see Fig.\ 5 of that paper which applies to $R = 40 $ Mpc
voids at $z_s =1$.  It also agrees to within a factor $\sim 2$ with
the predictions of weak lensing theory using an approximate power spectrum for our
void model, as discussed in Appendix \ref{appA}.

Figure \ref{fig:Convergence} shows how our estimated standard deviation
${\hat \sigma}_m(N)$ varies with number of runs $N$.  The quantity plotted is
$\log_{10}|{\hat \sigma}_m /
\sigma_{m}-1|$, where $\sigma_{m} = 0.03135$ is an
estimate of the $N \to \infty$ limit, here taken from our largest run
with $N = 10^6$.  This plot exhibits several interesting features that
are in good agreement with the analytical model described in Sec.\
\ref{sec:analyticalmodel}.  First, in the low $N$ regime at say $N
\sim 10^4$, the values of ${\hat \sigma}_m(N)$ differ systematically
from the asymptotic value by a few tens of percent, reflecting the
difference between $\sigma_{m,{\rm med}}$ and $\sigma_m$.  Second,
there is a somewhat smaller scatter in this regime, of $\sim 5\%$, in
agreement with the prediction (\ref{smallqci2}).  Third, there is a
transition to a different behavior at $N \sim 3 \times 10^5$, after
which both the scatter and systematic deviation from the asymptotic
value are much smaller.

In the rest of this paper, we will use the value $N = 10^6$ unless
otherwise specified.  From Fig. \ref{fig:Convergence} this corresponds to an accuracy
of $\sim 1$ percent.


%
\begin{figure}
\bec
\ifx\dofigures\undefined
\else
 {\includegraphics[scale=0.5]{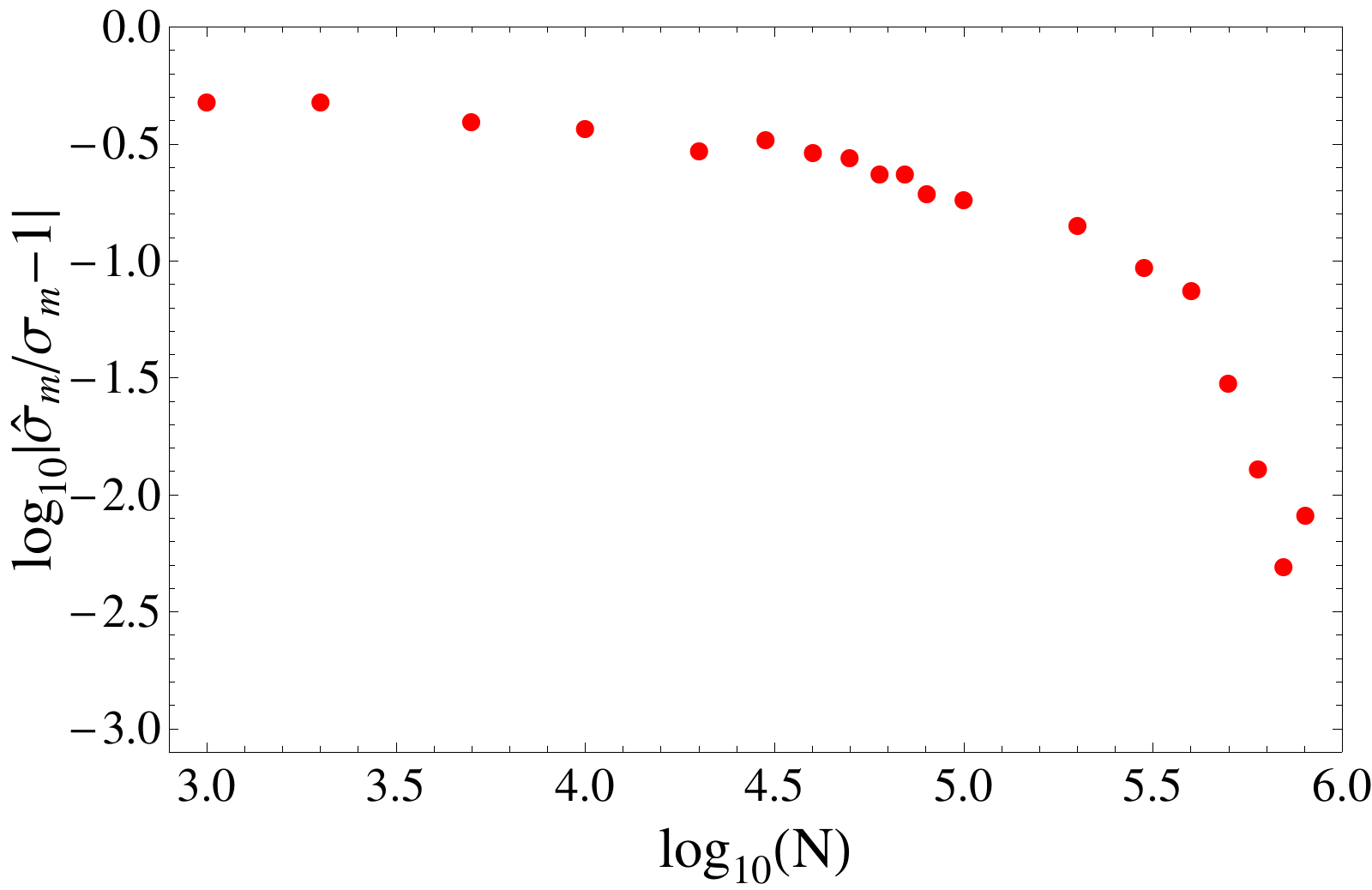}}
\fi
\caption{The estimator ${\hat \sigma}_m$ of
  the standard deviation of the distribution of
  magnitude shifts $\Delta m$, as a function of number $N$ of runs,
for sources at redshift $z_s=1$, comoving voids radius
$R = 35$ Mpc, and fraction of void mass on the shell today $f_0 =
0.9$.  The plotted quantity is $\log_{10}|{\hat \sigma}_m /
\sigma_m-1|$,
where $\sigma_m = 0.03135$ is an estimate of the $N \to
\infty$ limit, here taken from our largest run with $N = 10^6$.}
\label{fig:Convergence}
\eec
\end{figure}

\begin{figure}
\bec
\ifx\dofigures\undefined
\else
 \includegraphics[scale=0.6]{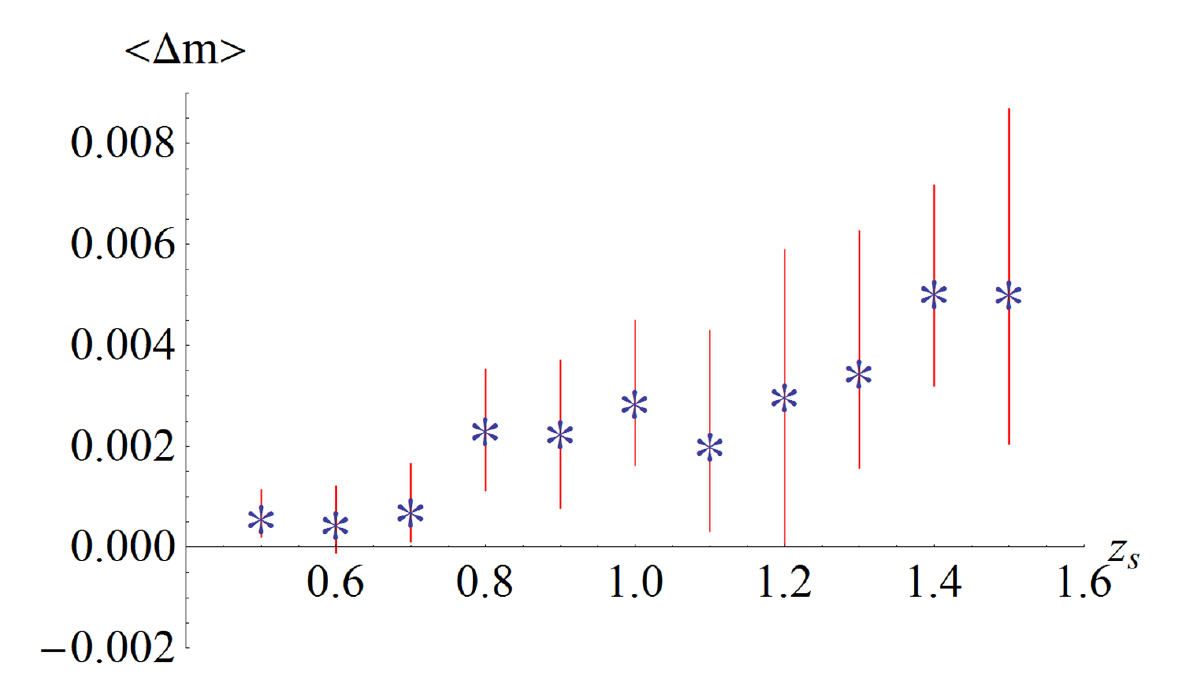}
 \includegraphics[scale=0.6]{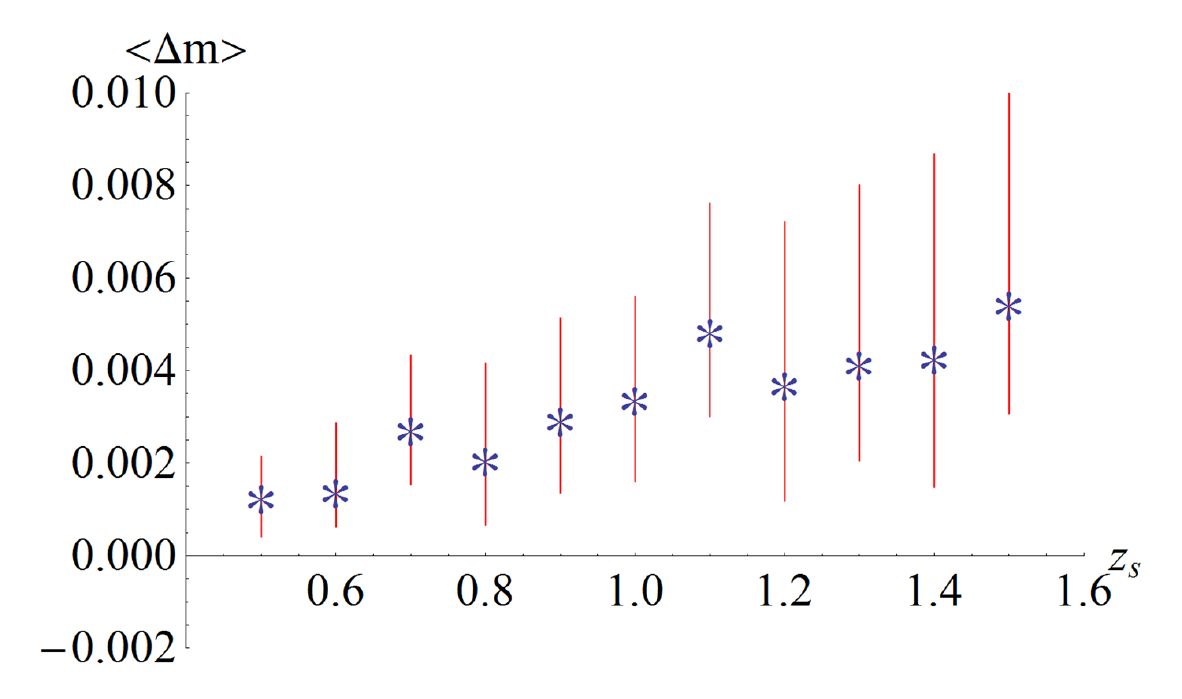}
\fi
\caption{[Top] The mean $\left< \Delta m \right>$ of the distribution of
  magnitude shifts $\Delta m$ as a function of source redshift $z_s$,
  for voids of radius $R = 35$ Mpc with fraction of mass on the shell
  today $f_0 = 0.9$, for $N = 10^6$ samples.  [Bottom] The same for
  $R = 100$ Mpc.}
\label{fig:Mean}
\eec
\end{figure}

We show in Fig.\
\ref{fig:Mean} the mean $\left< \Delta m \right>$ of the distribution
as a function of source redshift $z_s$, for $R = 35 $ Mpc and $N=
2\times 10^6$.  The errors shown are estimated by dividing the data into 200
groups of 10000 runs.  The effect of the nonzero mean on cosmological
studies
cannot be reduced by using a large number of supernovae, unlike the
effect of the dispersion $\sigma_m$.  However, the mean $\left< \Delta
  m \right> \sim 0.003$ magnitudes shown in Fig.\
\ref{fig:Mean} is too small to impact cosmological studies in the foreseeable
future.

\begin{figure}
\bec
\ifx\dofigures\undefined
\else
 {\includegraphics[scale=0.55]{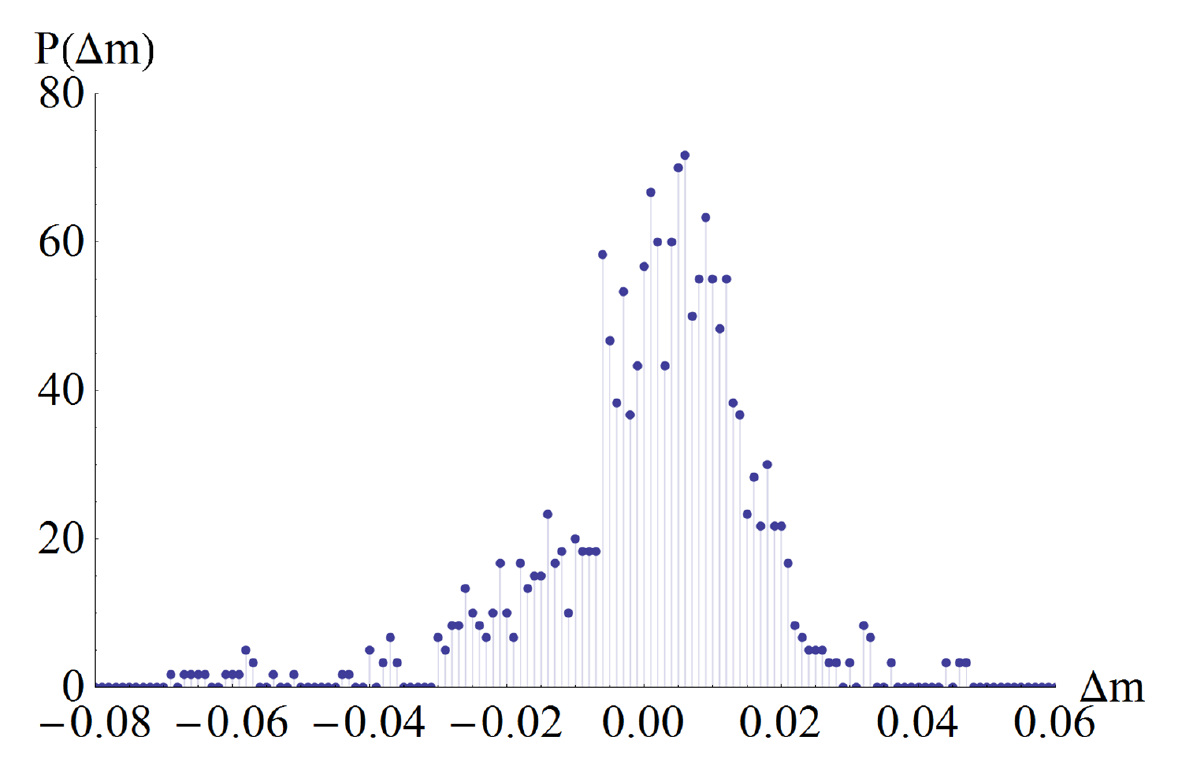}}
 {\includegraphics[scale=0.55]{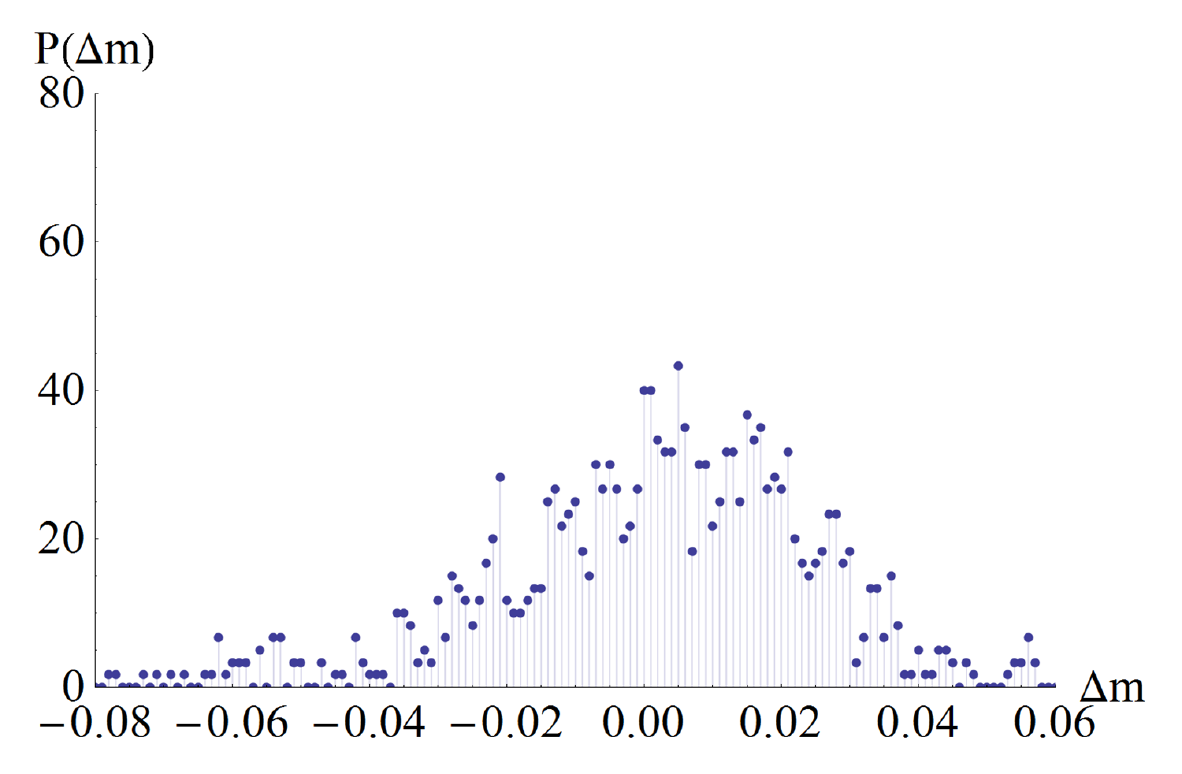}}
 {\includegraphics[scale=0.55]{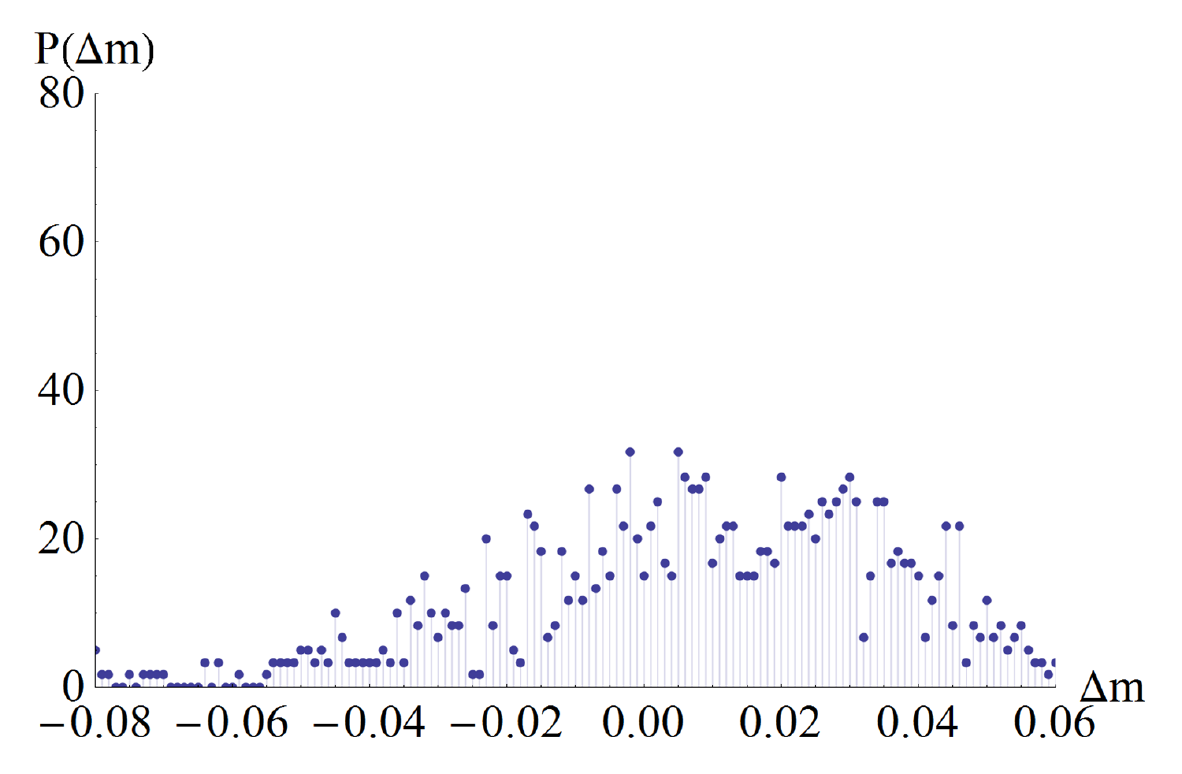}}
\fi
\caption{The probability distributions of magnitude shifts $\Delta m$
  for simulations with sources at redshifts of $z_s=1.1$ (top), $z_s=1.6$
  (middle) and
  $z_s=2.1$ (bottom), for comoving voids of radius $R=$35 Mpc
  with 90\% of the void mass on the shell today.}
\label{fig:PDF1}
\eec
\end{figure}

\begin{figure}
\bec
\ifx\dofigures\undefined
\else
 {\includegraphics[scale=0.55]{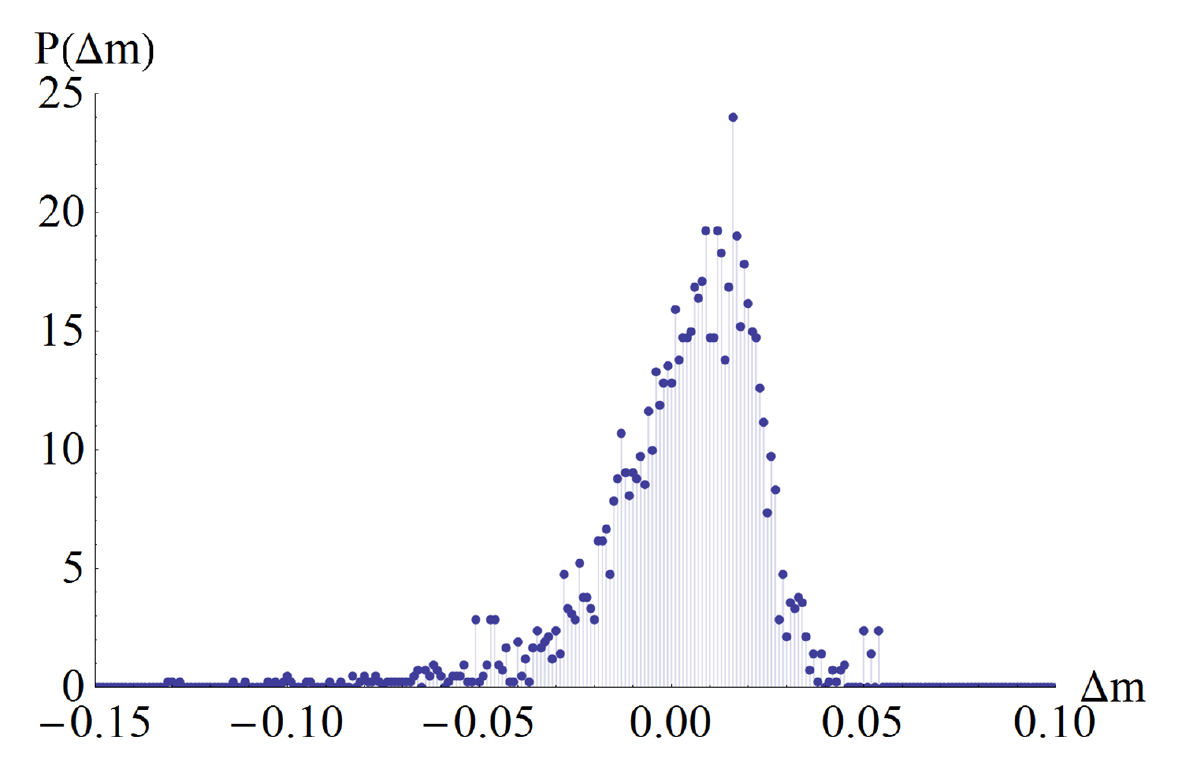}}
 {\includegraphics[scale=0.55]{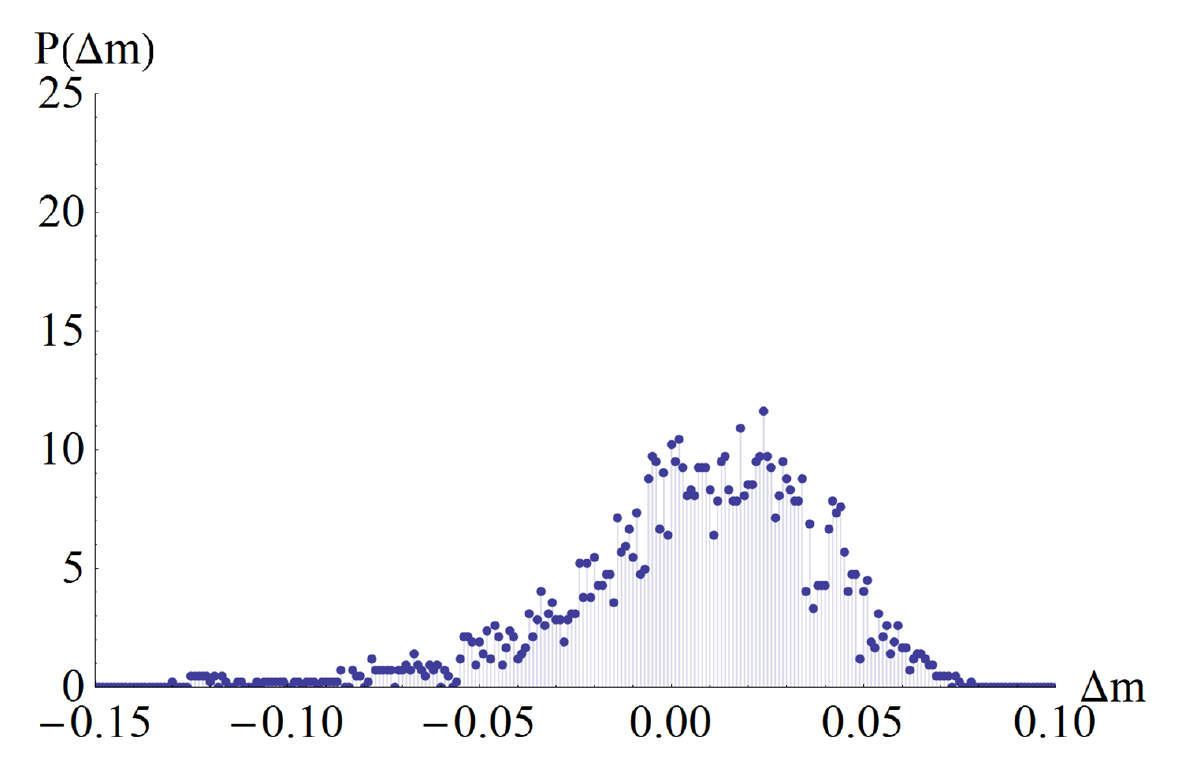}}
 {\includegraphics[scale=0.55]{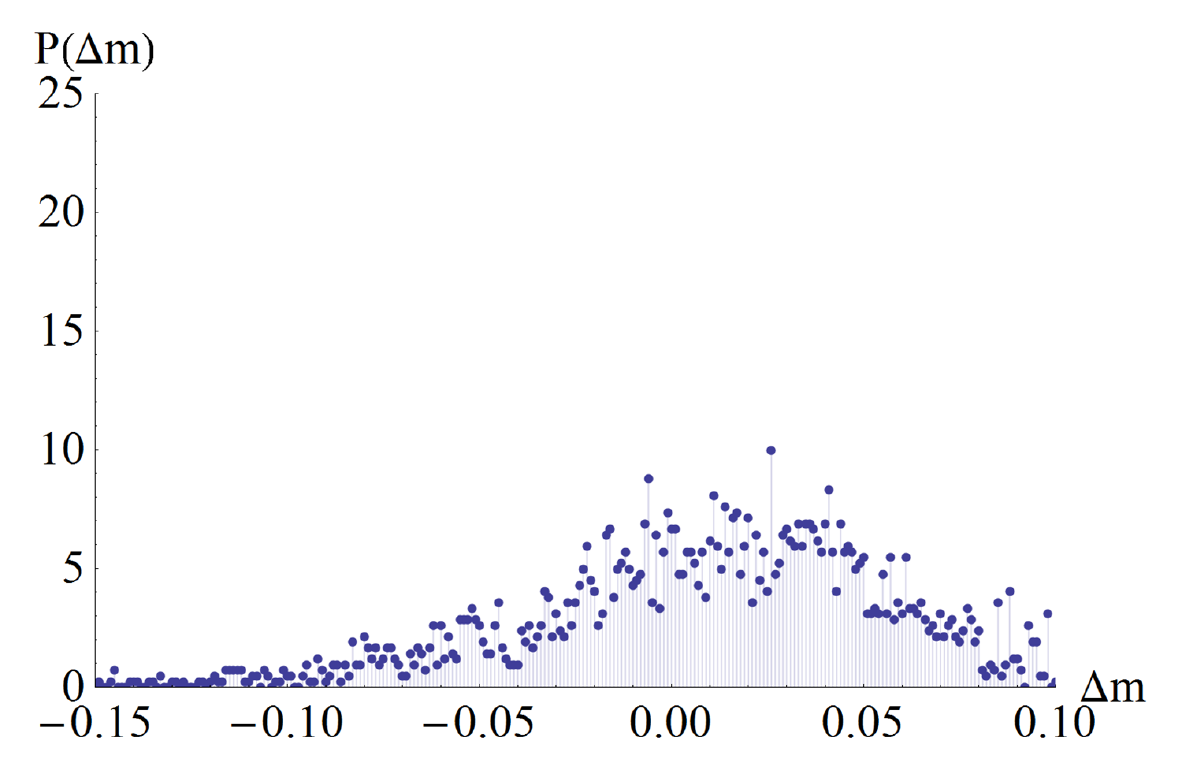}}
\fi
\caption{The probability distributions of magnitude shifts $\Delta m$,
at source redshifts $z_s$ of $1.1$ (top), $1.6$ (middle) and $2.1$
(bottom),
as in Fig.\ \protect{\ref{fig:PDF1}} except with comoving void radius of $R=100$
Mpc.}
\label{fig:PDF2}
\eec
\end{figure}

\begin{figure}
\bec
\ifx\dofigures\undefined
\else
 {\includegraphics[scale=0.55]{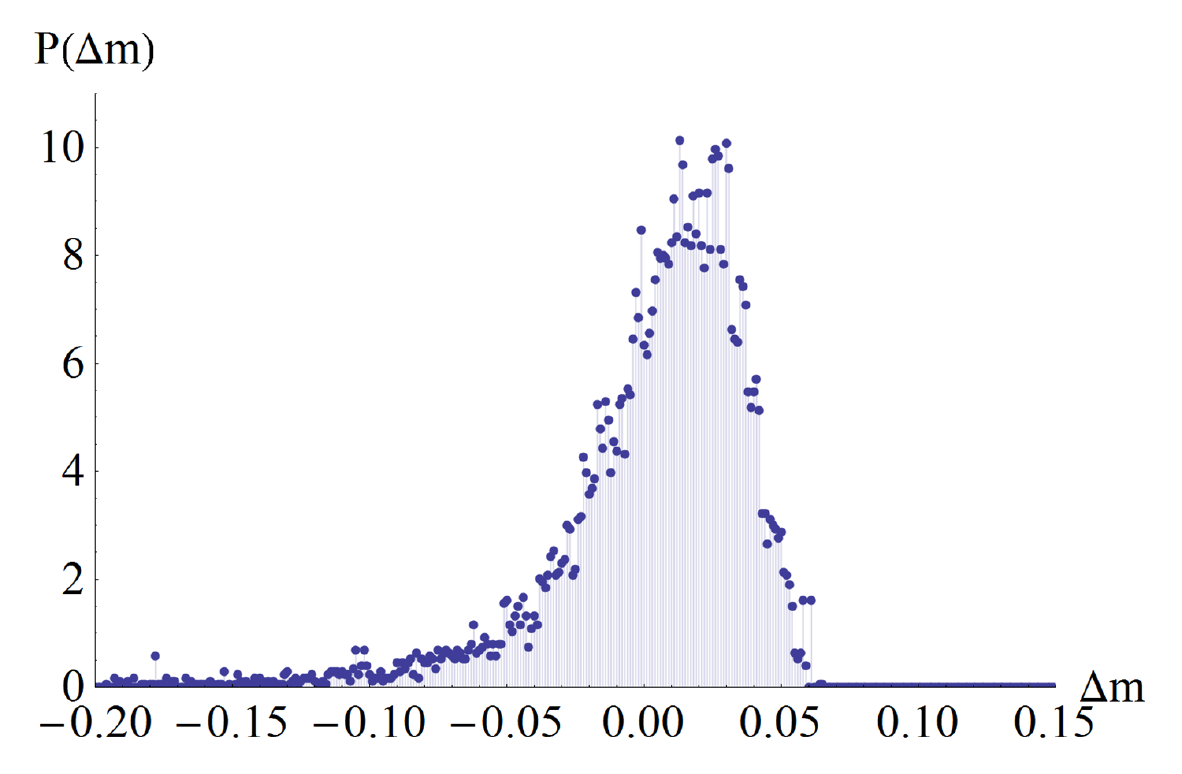}}
 {\includegraphics[scale=0.55]{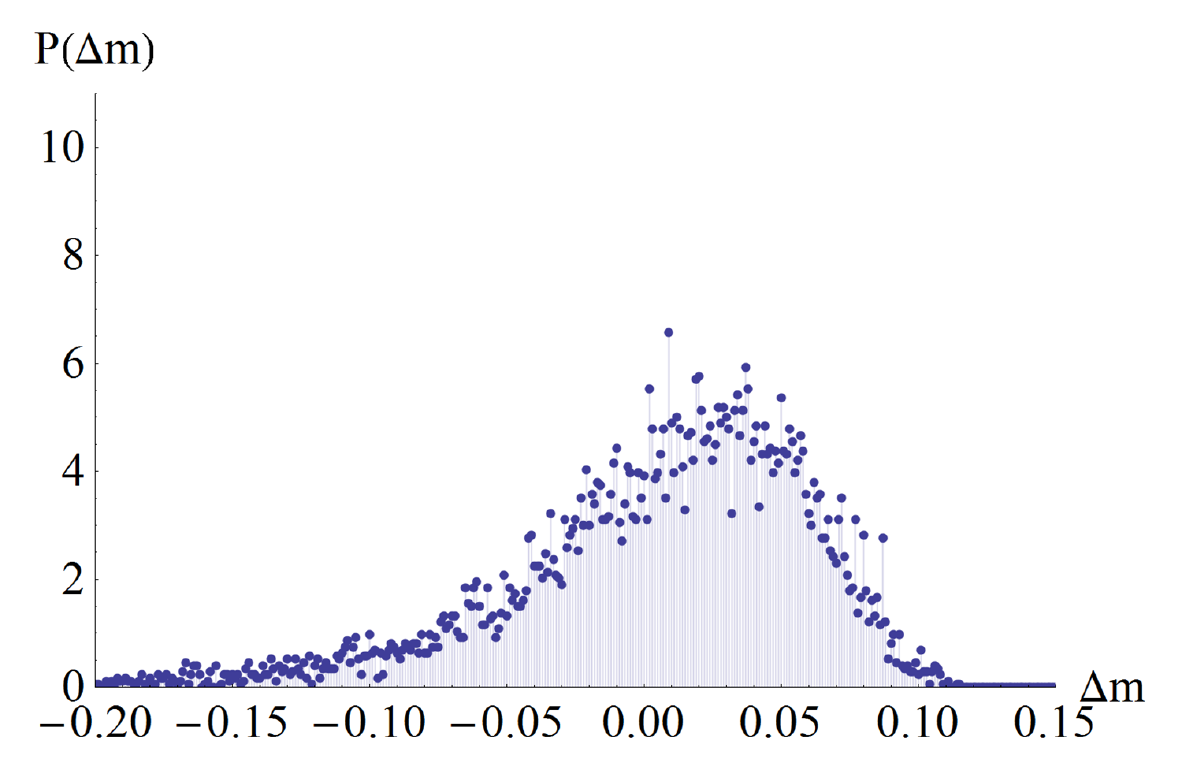}}
 {\includegraphics[scale=0.55]{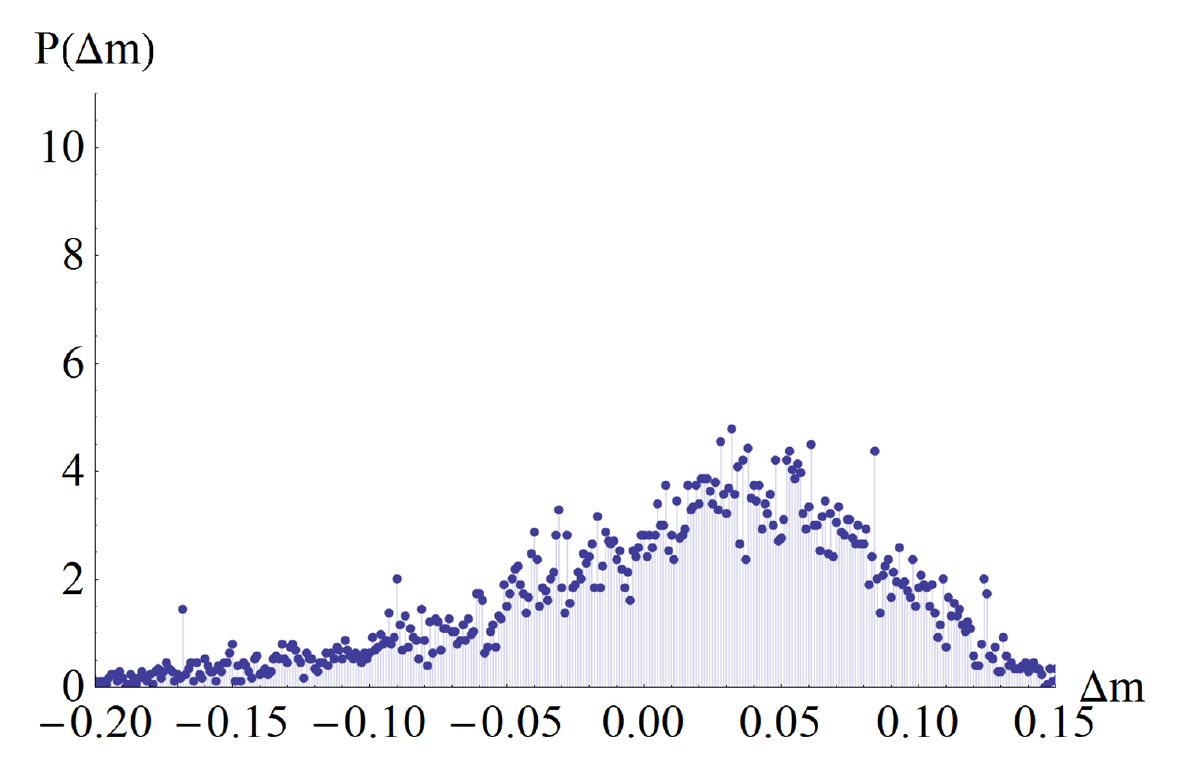}}
\fi
\caption{The probability distributions of magnitude shifts $\Delta m$,
at source redshifts $z_s$ of $1.1$ (top), $1.6$ (middle) and $2.1$
(bottom),
as in Fig.\ \protect{\ref{fig:PDF1}} except with comoving void radius of $R=350$
Mpc.}
\label{fig:PDF3}
\eec
\end{figure}

In Figs. \ref{fig:PDF1}, \ref{fig:PDF2}, and \ref{fig:PDF3} we show the
probability distributions of magnitude shifts $\Delta m$ for some other
cases: source redshifts of $z_s = 1.1,1.6$ and $2.1$, and void radii of
$R = 35$ Mpc, 100 Mpc, and 350 Mpc.
We now turn to a discussion of the dependence of our results on these
parameters, as well as on the fraction of mass in the shell today $f_0$.

\subsection{Dependence on void size}
\label{sec:voidsize}

\begin{figure}
\bec
\ifx\dofigures\undefined
\else
 \includegraphics[scale=0.65]{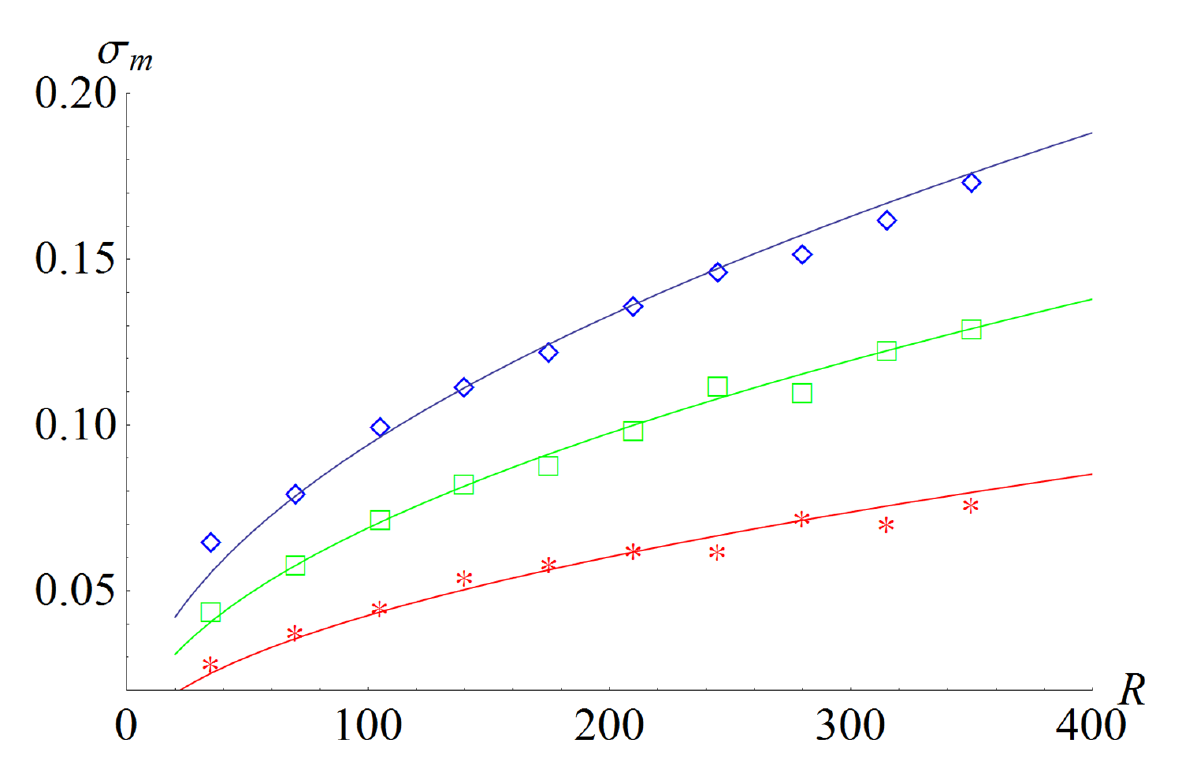}
\fi
\caption{The standard deviation $\sigma_m$ of the distribution of
  distance modulus shifts $\Delta m$ as a function of void radius $R$, computed
using $N = 10^6$ runs for each point.
The bottom line (stars)
is for sources at $z_s=1.1$, the middle line (squares) is $z_s=1.6$,
and the top line (diamonds) is $z_s=2.1$. Void radii range from
35 to 350 Mpc and the fraction of void mass on the shell today is $f_0 =
0.9$.  The lines
are fits of the form $\sigma_{m}\propto\sqrt{R}$.}
\label{fig:sigmam}
\eec
\end{figure}

In Fig.\ \ref{fig:sigmam} we show the standard deviation $\sigma_m$ of
the magnitude shift as a function of void size $R$, for three
different redshifts, $z_s = 1.1, 1.6, 2.1$.  To a good approximation
the standard deviation grows as the square root of the void size,
$\sigma_m \propto \sqrt{R}$.  We can understand this scaling by making
some order of magnitude estimates.

In making these estimates, we consider two different classes of rays.
Consider first rays that never come very close to the
shell of any of the voids, i.e. we exclude the case $b - R \ll R$,
where $b$ is the impact parameter.
The potential perturbation $\Delta \phi$ for passage through a void is
of order $\Delta \phi \sim f R^2 H_0^2$, where $f$ is the fraction of
void mass in the shell (or equivalently the fractional density
perturbation in the void interior).  The contribution to the lensing
convergence from this void is then of order $\kappa \sim \Delta \phi /
(H_0 R )
\sim f H_0 R$.  Next, the trajectory of rays is a random
walk, so the net lensing convergence is the rms convergence
for a single void multiplied by the square root of the number $\sim
1/(H_0 R)$ of voids.  Thus the contribution to the rms magnitude shift
from this class of rays is of order
\be
\sigma_m \sim f \sqrt{H_0 R}.
\label{eq:OE}
\ee

Consider next rays which just graze the shell of at least one of the voids.
These grazing rays are subject to large deflections,
because of the $\delta$-function in density on the surface of the
void.  The large deflections cause
cause the second moment $\left< \kappa^2 \right>$ of the
lensing convergence to diverge, as discussed in
Sec. \ref{sec:variance}.
However, the standard deviation of the
magnitude shift $\Delta m$ is still finite, because of the
logarithmic relation (\ref{eq:25b}) between $\Delta m$ and $\kappa$.

For estimating the effect of these grazing rays, we neglect shear.
The convergence $\kappa$ of the grazed void will be of order unity or
larger if the impact parameter $b$ is
$b=R(1-\varepsilon)$, where $\varepsilon \sim f^2 R^2 H_0^2$,
from Eq.\ (\ref{WLR}).  This will occur with probability $\sim
\varepsilon$.  The contribution of these rays to $\left< (\Delta m)^2
\right> \propto \left< [\ln(1-\kappa)]^2 \right>$ will be of order
$\varepsilon$ times the number $\sim 1/(H_0 R)$ of voids, or
$\sigma_m \sim f \sqrt{H_0 R}$, the same as the result (\ref{eq:OE})
for the non-grazing rays.

These considerations show that both the underdense void and the
mass-compensating shell make
substantial, comparably large contributions to $\sigma_m$. This
suggests that it may be important to refine the shell model to include
its fragmentation into localized overdensities
representing galaxy clusters and galaxies, as discussed in Sec.\
\ref{sec:finite} above.

\subsection{Dependence on fraction of void mass on the shell}
\label{sec:nonlin}

\begin{figure}
\bec
\ifx\dofigures\undefined
\else
\includegraphics[scale=0.5]{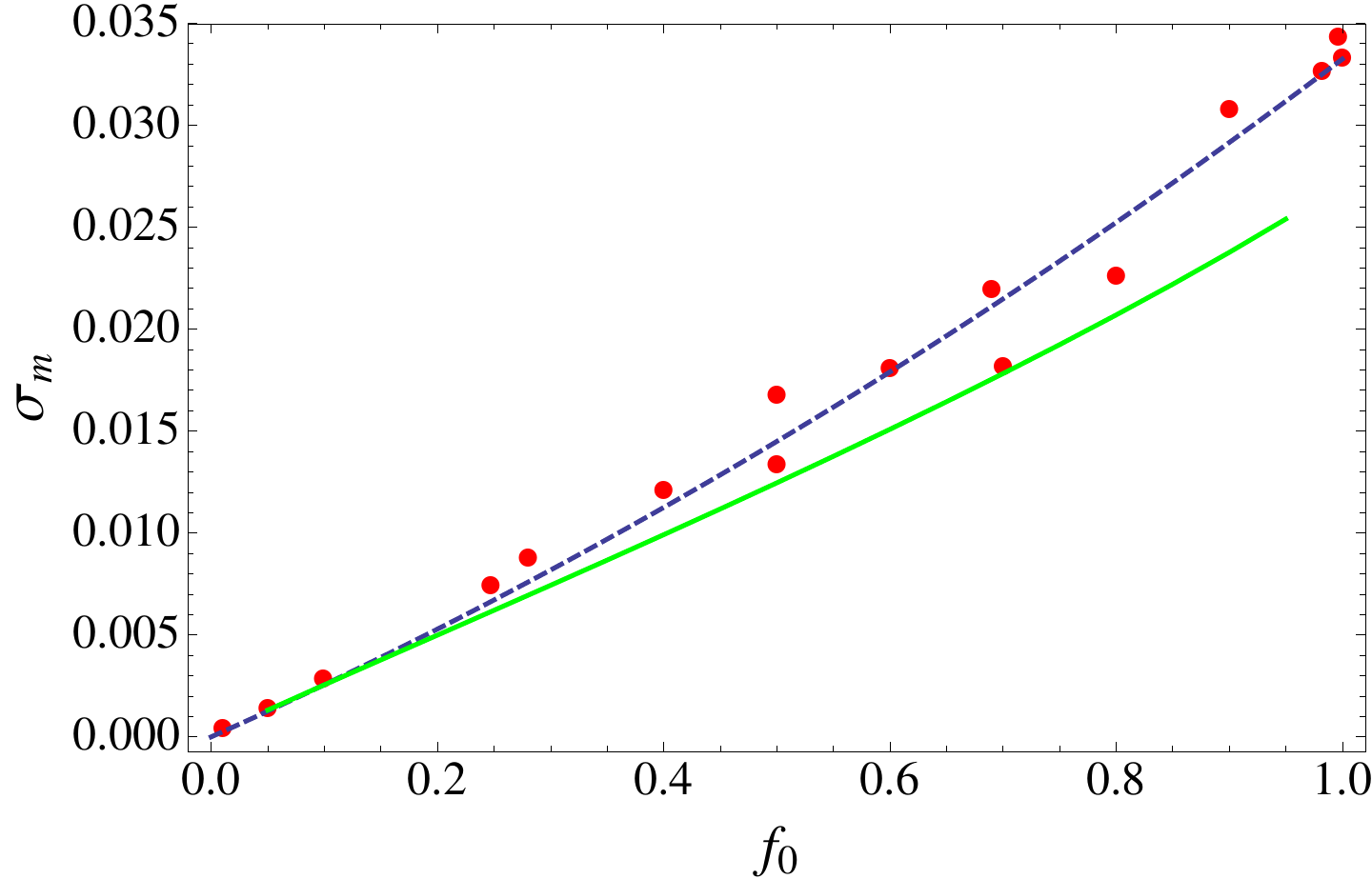}
\fi
\caption{The standard deviation $\sigma_m$ as a function of the
  fraction $f_0$ of the void mass on the shell today, for void radii of
  $R = 35$ Mpc and source redshift of $z_s=1$, computed using $N =
  10^6$ runs for each point.  The dashed blue curve is a fit of the
  form $\sigma_m = \alpha f_0 +
  \beta f_0^2$.  This plot shows that there are nonlinearities present
  at the level of $\sim 30-40\%$.  The solid green curve is the
  analytic model (\ref{analytic0}) -- (\ref{cutoff1}), which is
  accurate to $\sim 30\%$.
}
\label{fig:nonlinear}
\eec
\end{figure}

In this subsection we discuss the dependence of the magnification
distribution on the fraction $f_0$ of void mass on the shell today,
or, equivalently, on the fractional overdensity $\delta \rho/\rho$,
cf.\ Eq.\ (\ref{eq:11}) above.
Figure \ref{fig:nonlinear} shows the results of our simulations for
$\sigma_m$ as a function of $f_0$ for $N = 10^6$,
together with a fit of the form (\ref{eq:estimate})
\be
\sigma_m(f_0) = \alpha f_0 +\beta f_0^2
\label{eq:estimate}
\ee
for some constants $\alpha$ and $\beta$.
We find that $\alpha = 0.025 \pm 0.006$ and $\beta = 0.0085 \pm 0.0064$.
Thus, the data show a statistically
significant deviation from linear behavior, of the order of $\sim 30-40\%$.

We now discuss the various sources of nonlinearity that arise in the
computation.  We will consider three different types of effects.

First, in weak lensing theory, the magnification is a linear
function of the density perturbation.  Our computation includes some
nonlinear effects that go beyond weak lensing theory, specifically
lens-lens coupling (the fact that the deflection due to one lens
modifies the deflection caused by subsequent lenses) and shear (the
effect of the non-trace components of the matrices ${\cal R}^A_{\ \
  B}$ and ${\cal A}^A_{\ \ B}$).
To explore the magnitude of these effects, we performed Monte
Carlo simulations where we compute
the lensing convergence for each void and add these to obtain the
total lensing convergence (\ref{WLR0}),
and then compute $\Delta m$ from $\kappa$ using the exact nonlinear
relation (\ref{eq:25})
for zero shear.  The resulting value of $\sigma_m$ for
$f_0 = 0.9$, $z_s = 1$, $R = 35$ Mpc, $N = 10^6$ is
$\sigma_m = 0.0292$, about $7 \%$ smaller than the value $\sigma_m =
0.0314$ obtained by multiplying the $4 \times 4$ matrices. Thus, there
is a $\sim 7\%$ change from lens-lens coupling and shear.
For $R = 100$ Mpc, the change due to lens-lens coupling and shear is
$\sim 10\%$.
%
%
%
We also performed simulations where we kept just the trace part of the
matrix ${\cal R}_{AB}$, in order to exclude the effects of shear, but
included lens-lens couplings by computing $4 \times 4$ matrices for
each void and multiplying all these matrices.  In this case the
deviations of $\sigma_m$ from the full simulations
are $\sim 3 \%$ for $f_0 = 0.9$, $z_s = 1$, $R= 35 \, {\rm Mpc}$
and $\sim 6\%$ for $R = 100$ Mpc.  Thus, corrections due to shear
are of this order.

These nonlinearities due to lens-lens coupling and shear are
significantly smaller than the nonlinearity shown in Fig.\
\ref{fig:nonlinear}.  Thus other sources of nonlinearity must
dominate.  For the remainder of this subsection we will neglect
lens-lens coupling and shear, to simplify the discussion.

\begin{figure}
\bec
\ifx\dofigures\undefined
\else
\includegraphics[scale=0.5]{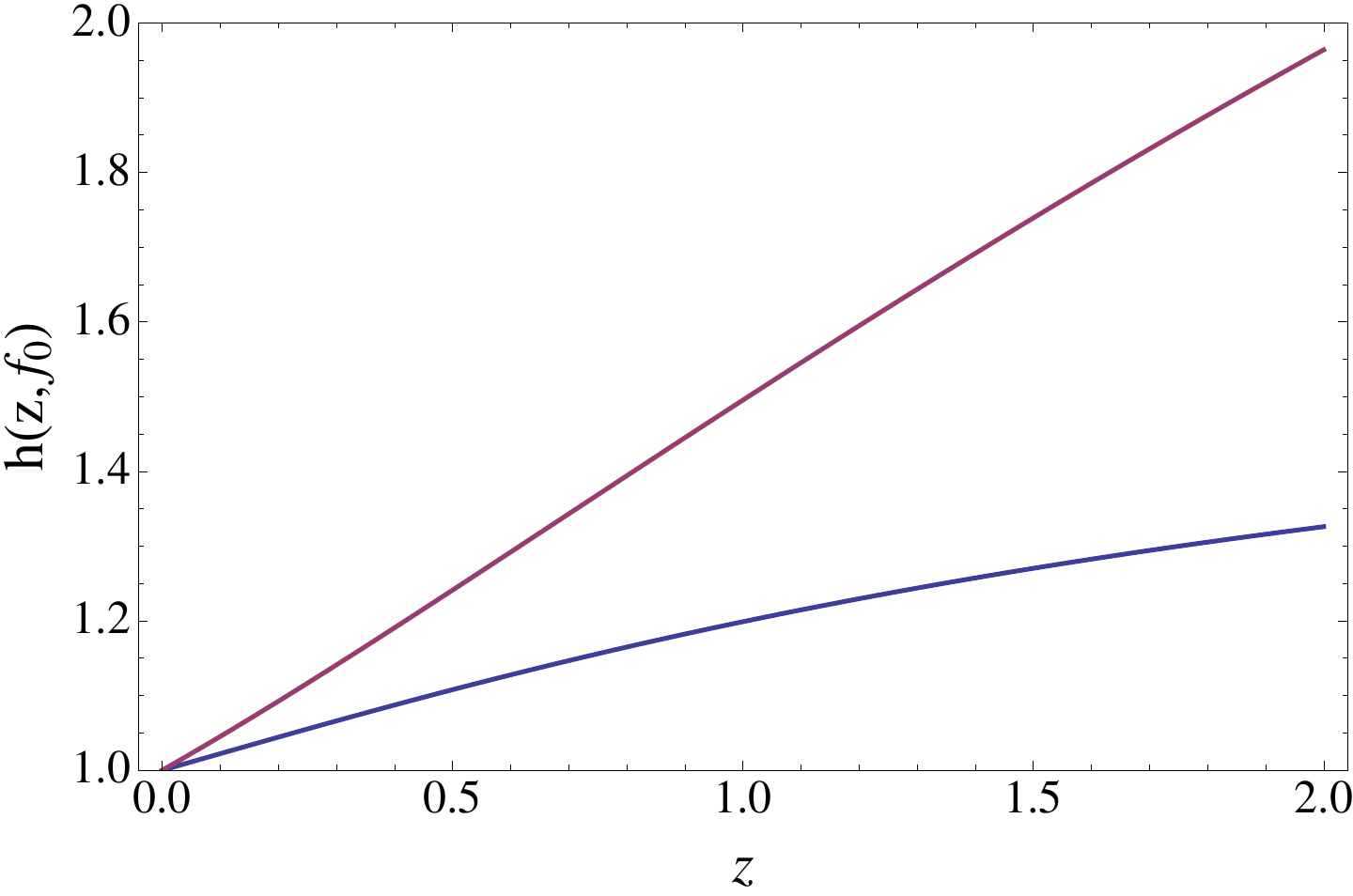}
\fi
\caption{The factor $h(z,f_0)$ by which nonlinear evolution corrects
the growth function $D_+(z)$ of linear perturbation theory, for our
void model.  The upper curve is for $f_0 = 0.9$ and the lower curve is
for $f_0 = 0.5$.}
\label{fig:nonlinear1}
\eec
\end{figure}

A second type of nonlinearity present in our computations is the fact
that the void mass fraction $f(z)$ at some redshift $z$ depends
nonlinearly on its value $f_0 = f(0)$ today, due to nonlinearity in
the void evolution.  Therefore, even if we make the weak-lensing
approximation of a linear dependence of the magnification on the
density perturbation $f(z)$, the magnification will still be a
nonlinear function of $f_0$.  We can parameterize this nonlinear evolution
effect by writing
\be
f(z;f_0) = f_0 D_+(z) h(z,f_0),
\ee
where $D_+(z)$ is the growth function of linear perturbation theory,
normalized so that $D_+(0) =1$, and
the function $h(z,f_0)$ incorporates the nonlinearity.  This function
satisfies $h(z,f_0) \to 1$ as $f_0 \to 0$ and also as $z \to 0$, and
can be computed using the results of Sec.\ \ref{sec:voidmodel} above.
Figure \ref{fig:nonlinear1} plots this function for $f_0  = 0.5$ and
$f_0 = 0.9$, and shows that the nonlinearities in the evolution are
significant.

\begin{figure}
\bec
\ifx\dofigures\undefined
\else
\includegraphics[scale=0.5]{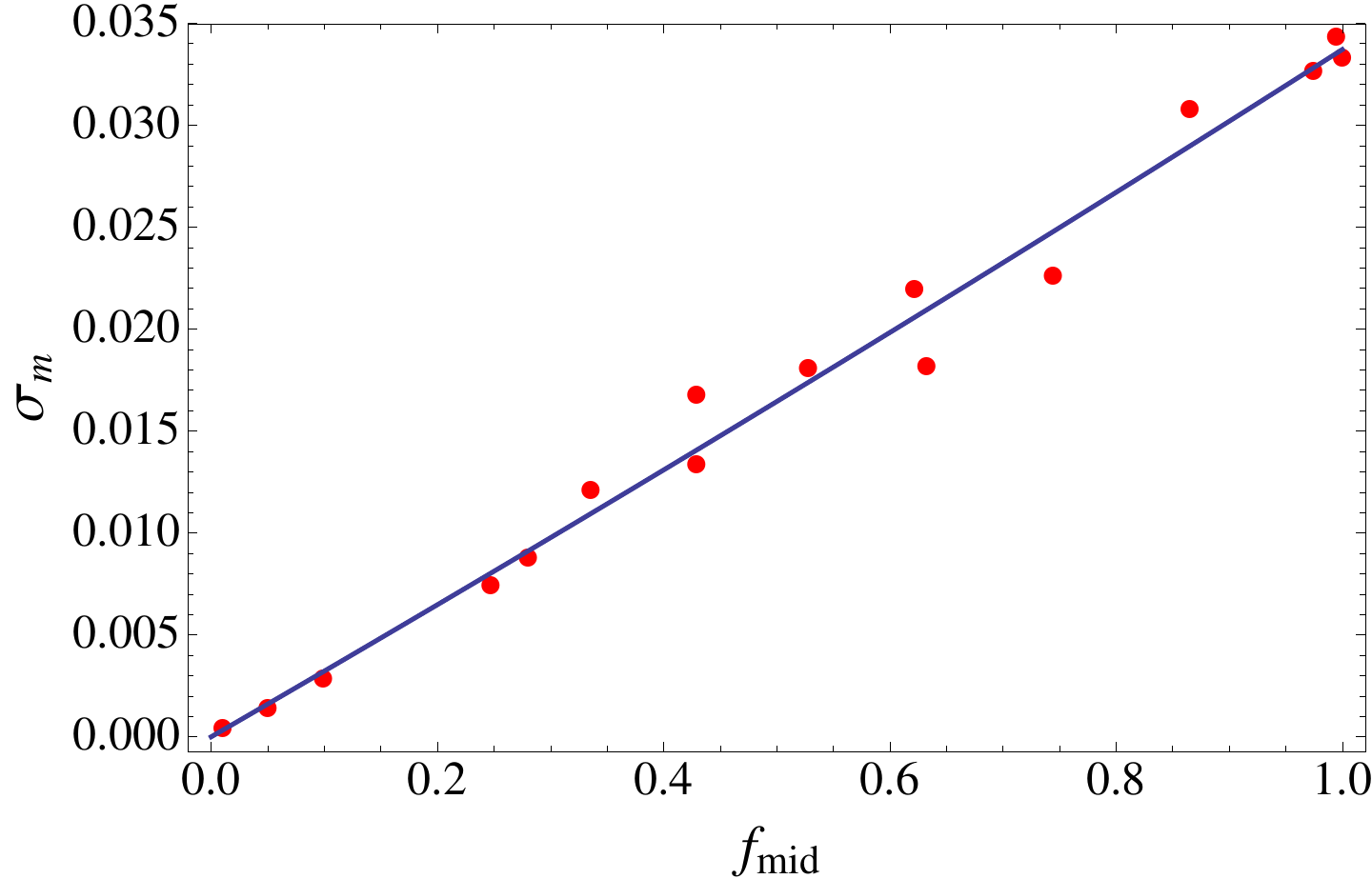}
\fi
\caption{The standard deviation $\sigma_m$ as a function of the
  fraction $f_{\rm mid}$ of the void mass on the shell for voids
  halfway to the source, for void radii of
  $R = 35$ Mpc and source redshift of $z_s=1$.
The solid line is a fit of the form $\sigma_m = \alpha f_{\rm mid} +
  \beta f_{\rm mid}^2$.  For this choice of parameterization there is no
  statistically significant nonlinearity detectable in the data.}
\label{fig:midpoint}
\eec
\end{figure}

This nonlinear evolution effect is the dominant source of nonlinearity
in our simulations.  To illustrate this, we define, for a given source
redshift $z_s$, the parameter
\be
f_{\rm mid} \equiv f(z_s/2,f_0).
\ee
In other words, $f_{\rm mid}$ is the fraction of void mass on the shell
for voids halfway to the source, the distance where most of the lensing occurs.
We can use $f_{\rm mid}$ instead of $f_0$ as a parameter to describe our
voids.  With this choice of parameterization, the nonlinear evolution
effect is significantly reduced.  This is illustrated in Fig.\
\ref{fig:midpoint}, which shows the same data as in Fig.\
\ref{fig:nonlinear}, but as a function of $f_{\rm mid}$ rather than
$f_0$.  The best fit parameters in the quadratic fit $\sigma_m =
\alpha f_{\rm mid} + \beta f_{\rm mid}^2$ are now
$\alpha = 0.032 \pm 0.005$, $\beta = 0.0016 \pm 0.0057$, showing that
there is no statistically significant nonlinearity.

A third type of nonlinearity in our simulations arises from the
nonlinear relation between the lensing convergence $\kappa$ and the
magnitude shift $\Delta m$.  This effect should be present in our data
but is quite small.  
If we neglect lens-lens coupling, shear, and the nonlinear evolution
effect, then we expect logarithmic terms in the relation between
$\sigma_m$ and $f_0$, of the form
\be
\sigma_m^2 \sim \alpha f_0^2 + \beta f_0^2 \ln f_0 + \ldots,
\label{logterms}
\ee
where $\alpha$ and $\beta$ are constants which are independent of
$f_0$.
This follows from the analysis of Sec.\ \ref{sec:variance} above,
where the logarithmic divergence in the variance is cutoff at $\kappa
\sim 1$; see Eqs.\ (\ref{analytic0}) and (\ref{cutoff1}). However our
data show that the logarithmic terms in Eq.\ (\ref{logterms}) are
quite small.

\begin{figure}
\bec
\ifx\dofigures\undefined
\else
\includegraphics[scale=0.6]{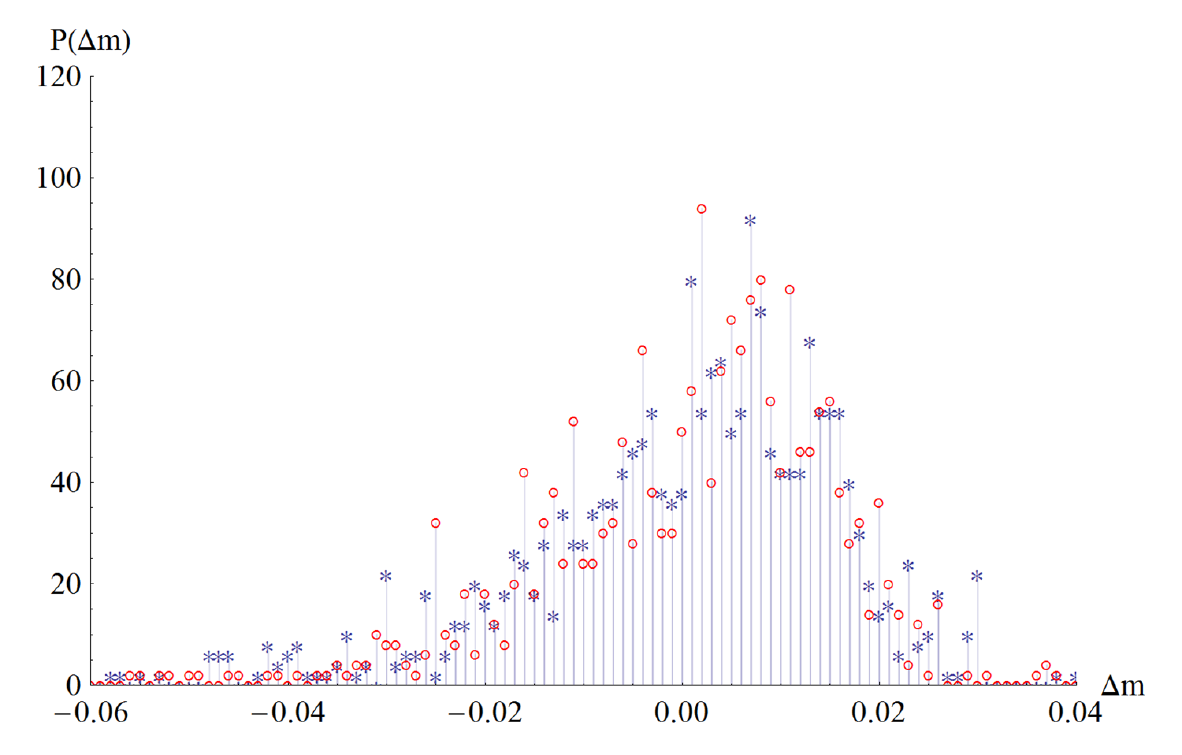}
\fi
\caption{A comparison of the probability distributions of magnitude
shifts $\Delta m$ in two different cases: fraction of mass on the
shell today fixed at $f_0 =0.9$ (circles), and $f_0$ drawn from a
distribution as described in the text (stars).  In both cases void
radius is $R = 35$ and source redshift is $z_s = 1.0$.  The spread in
the shell surface densities gives
rise to a wider distribution of magnitude shifts, by about $\sim 3\%$.}
\label{fig:PDFcomparison}
\eec
\end{figure}

Next, we discuss
the effects of allowing a distribution of
values of void mass fraction on the shell $f_0$ in our simulations,
rather than having a fixed value.
We performed simulations where
we pick a value of $f$ for
for each void crossing according to the following prescription.
We choose a random values for $1/a_{0}$ from a Gaussian
distribution with a mean of $8$ and a variance of $30$, truncated
to lie in the range that corresponds to $0\leq f\leq1$.
Figure \ \ref{fig:PDFcomparison} compares the probability
distributions for magnitude shifts with and without variations in $f$.
Treating $f$ as a random variable increases the standard deviation
$\sigma_m$ by $\sim 3\%$.

\subsection{Dependence on source redshift}

\begin{figure}
\bec
\ifx\dofigures\undefined
\else
 \includegraphics[scale=0.65]{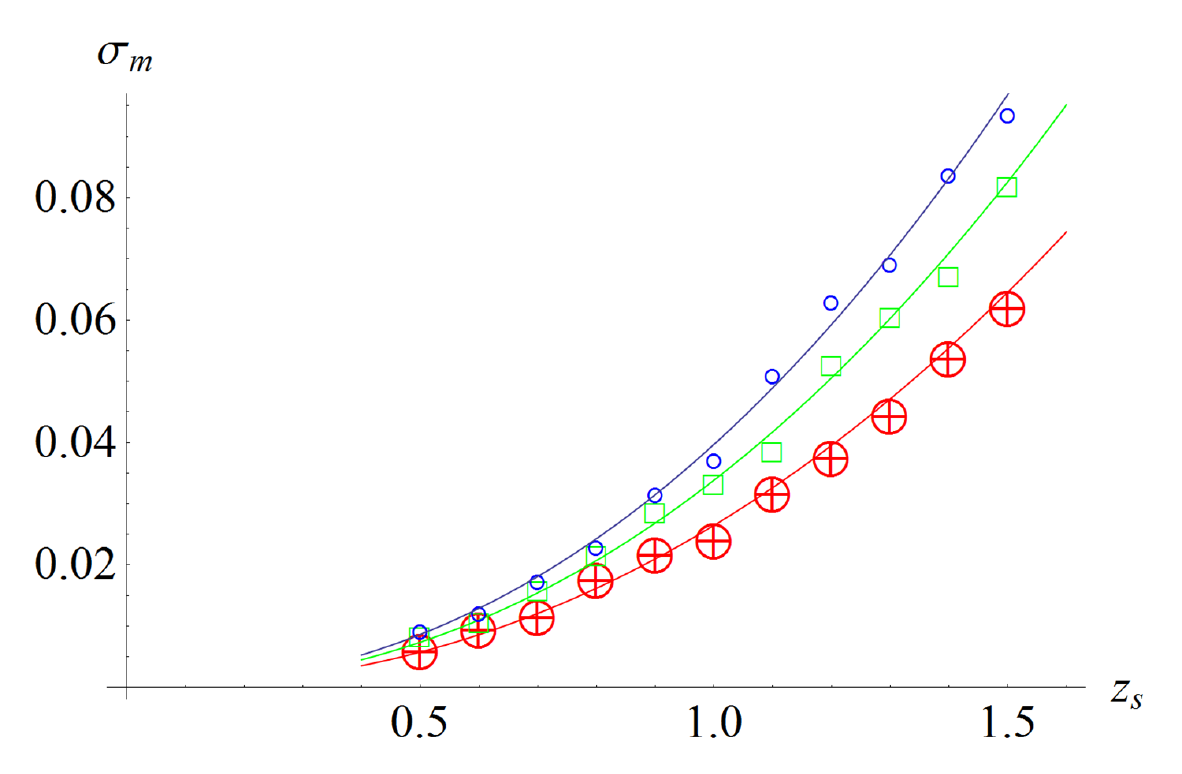}
\fi
\caption{The standard deviation $\sigma_m$ as a function of source
  redshift $z_s$, computed using $N = 10^6$ runs, for voids of radii
  $R = 35$ Mpc (red, crossed circles),
$70$ Mpc (green, squares), and $105$ Mpc (blue, circles).  The lines
are fits proportional to the analytic
estimate (\protect{\ref{eq:45-1}}).}
\label{fig:sigmam1}
\eec
\end{figure}

Figure \ref{fig:sigmam1} shows the standard deviation $\sigma_m$ of
the magnitude shift distribution as a function of source redshift
$z_s$, for three different void sizes.  The standard deviation
increases with redshift faster than $z_s$.  This increase is due in
part to the increasing number of voids but there are additional factors.

To understand the redshift dependence analytically we use the expression
for the dispersion in lensing convergence from weak lensing
theory, given by Eq.\ (\ref{eq:A1}) in Appendix \ref{appA}.
The matter power spectrum $\Delta(k,z)^2$ for our void model is
proportional to $f(z)^2$, so we obtain that
\be
\left< \kappa^2 \right> \propto \int_0^{x_s} dx w(x,x_s)^2 f(z)^2,
\label{eq:45-1}
\ee
where $w(x,x_s) = (1+z) H_0 x (x_s-x)/x_s$ and $f(z)$ is defined by
Eq.\ (\ref{eq:8a}).
In the range of redshifts $0.5 \le z_s \le 1.5$ this redshift
dependence is approximately a power law, proportional to $z_s^{1.35}$,
to within $\sim 5\%$ percent\footnote{The asymptotic behavior at large $z_s$
  is that the expression (\protect{\ref{eq:45-1}})
  increases linearly in $z_s$.}.  This redshift dependence agrees with the
results of our simulations shown in Fig.\ \ref{fig:sigmam1} to within
$\sim 10\%$.

\subsection{Numerical fit to parameter dependence}

We complete this part of the analysis by giving a three
parameter fit for the standard deviation $\sigma_m$ as a function of void radius
$R$, fraction of void mass on the shell today $f_0$, and source redshift
$z_s$.  The result is
\be
\sigma_m
\approx
(0.027 \pm 0.0007)
\left( \frac{R}{35 \, {\rm Mpc}}  \right)^\alpha
\left( \frac{f_0}{0.9} \right)^\beta \left( \frac{z_s}{1.0} \right)^\gamma,
\ee
where the parameters are $\alpha = 0.51 \pm 0.03$, $\beta = 1.07
\pm
0.04$, $\gamma = 1.34 \pm 0.05$.
This fit is accurate to $\sim 20 \%$
for $35 \, {\rm Mpc} \, \le R \le 350 \, {\rm Mpc}$, $0.01 \le f_0 \le 0.9$, $0.5 \le z_s \le 2.1$.

\section{Bias due to sources occurring preferentially in high density regions}
\label{sec:bias}

For sources which are randomly distributed in space, it is known that the total
expected apparent luminosity of a source, including all primary and secondary images,
must agree with that of the background FRW model \cite{HW}.  Hence, in situations
where the probability of caustics can be neglected, the probability
distribution (\ref{OR}) of magnifications $\mu$ must be {\it
  unbiased}.  Biases arise in our computations because of caustic
effects, and also because we study the probability distribution of the
magnitude shift $\Delta m$, which is a nonlinear function
of $\mu$, cf.\ Eq.\ (\ref{eq:25a}).

However, there is an additional fundamental source of bias which
arises from the fact that sources are {\it not} randomly distributed
in space, and instead preferentially occur in high density
regions, where they are more likely to be close to a lens.  This is
the source-lens clustering effect \cite{SLC}.  In this section, we
make an analytical estimate of the bias $\delta m$ of the distribution of magnitude
shifts that is due to source-lens clustering in our void model.

In our computations so far in this paper, we have placed the source
outside the voids, in the FRW regions.
However, in reality most matter is concentrated on the edges of
voids, and so sources are more likely to be on the void edges.
If we demand that sources always be located on void edges, then
the mean of the distribution is shifted by an amount (see derivation below)
\begin{equation}
\delta
m=\frac{1}{3\ln\left(10\right)}\left(1+z_{s}\right)H_{0}^{2}R^{2}\Omega_{M}f_{s}.
\label{eq:46-1}
\end{equation}
Here $z_s$ is the redshift of the source and $f_{s} = f(z_s)$ is the fraction
of mass on the shell for voids at the source redshift. Evaluating
this estimate for $\Omega_{M}=0.3$, $z_s=1.0$, $R=35$ Mpc, $f_0=0.9$
gives $\delta m\sim5 \times 10^{-6}$, and $\delta m \sim 5 \times
10^{-4}$ for $R = 350 \, {\rm Mpc}$.  These biases are
below the accuracy of upcoming cosmology surveys.

Turn now to the derivation of the formula (\ref{eq:46-1}). We start from the
standard formula (\ref{eq:42-1}) for the lensing convergence in weak lensing theory.  We consider just the contribution to $\kappa$ from
the last void. In the integral, over this void, we approximate the
factors $x$ and $1/a_{\rm ex}\left(z\right)$ as constants. Writing $\eta=x_{s}-x$
we obtain\begin{equation}
\kappa_{{\rm last}\: {\rm
    void}}=\frac{3}{2}H_{0}^{2}\left(1+z_{s}\right)\Omega_{M}\int_{{\rm
    last}\: {\rm void}}\eta\delta_{\rm m}({\bf x},t)d\eta.\label{eq:47}\end{equation}
We also neglect the time dependence of $\delta_{\rm m}({\bf x},t)$
for integrating over the last void.

We now consider two different models for randomizing the relative
displacement between the center of the last void and the source. We
denote by $b$ the transverse displacement of the void center from
the line of sight, as before, and denote by $\eta_{v}$ the distance
from the void center to the plane through the source perpendicular
to the line of sight.

In our first model, we assume $b$ and $\eta_{v}$ are randomly distributed,
proportional to $bd\eta_{v}db$, with $0\leq\eta_{v}\leq R$ and $0\leq b\leq R$.
Computing the integral (\ref{eq:47}) for our void model (\ref{eq:11})
gives
\bea
\label{jjj}
\kappa_{{\rm last}\: {\rm
    void}}&=&\frac{3}{2}H_{0}^{2}\Omega_{M}\left(1+z_{s}\right)
 \\
&& \times
\left\{ \begin{array}{c}
-2f_{s}\eta_{v}\alpha+\frac{2fR^{2}\eta_{v}}{3\alpha} \nonumber \\
-\frac{1}{2}f\left(\eta_{v}+\alpha\right)^{2}+\frac{fR^{2}}{3\alpha}\left(\eta_{v}+\alpha\right)\end{array}\right.\begin{array}{c}
\eta_{v}>\alpha \nonumber \\
\eta_{v}<\alpha\end{array}
\label{eq:48}
\eea
where $\alpha=\sqrt{R^{2}-b^{2}}$. Now averaging over $b$ and $\eta_{v}$
gives the expected value of $\left\langle \kappa_{{\rm last}\: {\rm void}}\right\rangle =\left(1+z_{s}\right)H_{0}^{2}R^{2}\Omega_{M}f_{s}/15$.

In the second model, we assume that $b$ and $\eta_{v}$ are correlated
so that the source is always on the surface of the void. The average
of $\kappa_{{\rm last}\: {\rm void}}\left(b,\:\eta_{v}\right)$ in this model
is
\begin{equation}
\left\langle \kappa_{{\rm last}\: {\rm void}}\right\rangle
=\int_{0}^{\pi/2}\sin\theta \, \kappa\left(R\sin\theta,\:
  R\cos\theta\right)d\theta,
\label{eq:49}
\end{equation}
which using the formula (\ref{jjj}) gives zero. Subtracting the means of
the two models gives an estimate of the bias, and multiplying the
result by $5/\ln10$ to convert from $\delta\kappa$ to $\delta m$
gives the formula (\ref{eq:46-1}).

\section{Conclusions}
\label{sec:conclusions}

In this paper, we presented a simple model to study the effects of
voids on distance modulus shifts due to gravitational lensing. A number
of future surveys will gather data on luminosity distances to various
different astronomical sources, to use them to constrain properties
of the source of cosmic acceleration.
The accuracy of the resulting constraints will be
degraded somewhat by lensing due to nonlinear large scale structures.
We studied this effect by considering a $\Lambda$CDM Swiss cheese cosmology with
mass compensating, randomly located voids with uniform interiors
surrounded by thin shells.

We used an algorithm to compute the probability distributions of distance
modulus shifts similar to that of Holz \&
Wald \cite{HW}. The rms magnitude shift
due to gravitational lensing of voids is fairly small; the dispersion
$\sigma_{m}$ due to 35 Mpc voids for sources at $z_s=1$ is $\sigma_{m}=0.031$,
which is $\sim 2 -3$ times smaller than that due to galaxy clusters
(see Appendix \ref{appA} below).
Also
the mean magnitude shift due to voids is of order $\delta m \sim
0.003\pm0.001$.
We also studied the bias that arises from the source-lens clustering effect,
and estimated that the contribution from voids to this bias is quite
small, of order $\delta m \sim 5 \times 10^{-6}$.
Refining our model by giving each void shell a finite thickness of
$\sim 1 $ Mpc reduces the dispersion $\sigma_m$ by a factor $\sim 2$.

We used our model to estimate the sizes of various nonlinear effects
that go beyond linear, weak-lensing theory.  We estimate that
for $R = 35$ Mpc the
dispersion $\sigma_m$ is altered by $\sim 4\%$ by lens-lens coupling,
by $\sim 3 \%$ by shear.
For 100 Mpc voids these numbers become $3\%$ and $6\%$
respectively.

Our simple and easily tunable model for void lensing can be used as
a starting point to study more complicated effects.
For example, one can use various algorithms to generate realizations
of distributions of non-overlapping spheres in three dimensional
space.  Given such a realization one could use the algorithm of this
paper to study correlations between magnifications along rays with
small angular separations, which would be relevant to future pencil
beam surveys \cite{pencil}.
Finally, our model is complementary to other simplified lensing models
in the literature that focus on lensing due to halos but neglect
larger scale structures, for example the model of Refs.\ \cite{KM09,KM11}.

\begin{acknowledgments}
This research was supported at Cornell by NSF grants PHY-0757735,
PHY-0555216, and PHY-0968820
and by NASA grant NNX 08AH27G.
RAV acknowledges support from the Kavli Institute for Cosmological
Physics at the University of Chicago through grants NSF PHY-0114422
and NSF PHY-0551142 and an endowment from the Kavli Foundation and
its founder Fred Kavli.
\end{acknowledgments}

\appendix

\section{Comparison with weak lensing theory }
\label{appA}

In this appendix we show that our results agree moderately well with
the predictions of weak lensing theory, by computing an approximate matter power
spectrum for our void model.  We also obtain an independent estimate
of the lensing due to voids by using the power spectrum of the
Millennium simulation\cite{Millennium}.

It is somewhat complicated to compute an exact power spectrum for our
distribution of voids.  As a simple model, we choose a two-void
probability distribution function for which the locations of the two
voids are independently and uniformly distributed inside some large
finite volume, except that the probability is set to zero when the
distance between the void centers is less than $2 R$. For this model,
using the void density profile (\ref{eq:11}), we find for the power
spectrum\footnote{This model is not completely consistent, since the
  power spectrum can become negative for large packing fractions.  The
  inconsistency is presumably a signal that our assumed 2-void probability
  distribution cannot be obtained starting from any symmetric
  non-overlapping $n$-void
  probability
  distribution.  We ignore this inconsistency here since the
  correlation effects that give rise to the correction factor in square
  brackets in Eq.\ (\protect{\ref{Delta2US}}) give only a small ($<1\%$)
  correction to $\left< \kappa^2\right>$ in any case.}
\be
\Delta(k,z)^2 = \frac{2 \alpha}{3 \pi} f(z)^2 k^3 R^3 j_2(k R)^2
\left[ 1 - 12 \alpha \frac{j_1(2 k R)}{kR} \right].
\label{Delta2US}
\ee
Here $\alpha$ is the void packing fraction, which is $\pi/6$ in our
model, $k$ is wavenumber, $j_1$ and $j_2$ are spherical Bessel
functions of the first kind, and $f(z)$ is
the fraction of the void mass in the shell, which can be computed as a
function of redshift using the results of Sec.\ \ref{sec:voidmodel}.
We note that this power spectrum is not an exact representation of our
void model, because in our procedure we first choose a direction to
the source and then generate a density perturbation field that depends on
this direction.  Thus, our procedure does not correspond exactly
to choosing a direction randomly in a pre-existing homogeneous,
isotropic random process\footnote{If the model were exactly homogeneous there
  would be a nonzero probability for the observer to be located inside
  a void.}, i.e. $\langle \delta \rho({\bf x}) \delta \rho({\bf y})
\rangle$ is not just a function of $|{\bf x} - {\bf y}|$.
Homogeneity is necessary in order to represent the two point
function in terms of a power spectrum.

\begin{figure}
\ifx\dofigures\undefined
\else
\includegraphics[scale=0.5]{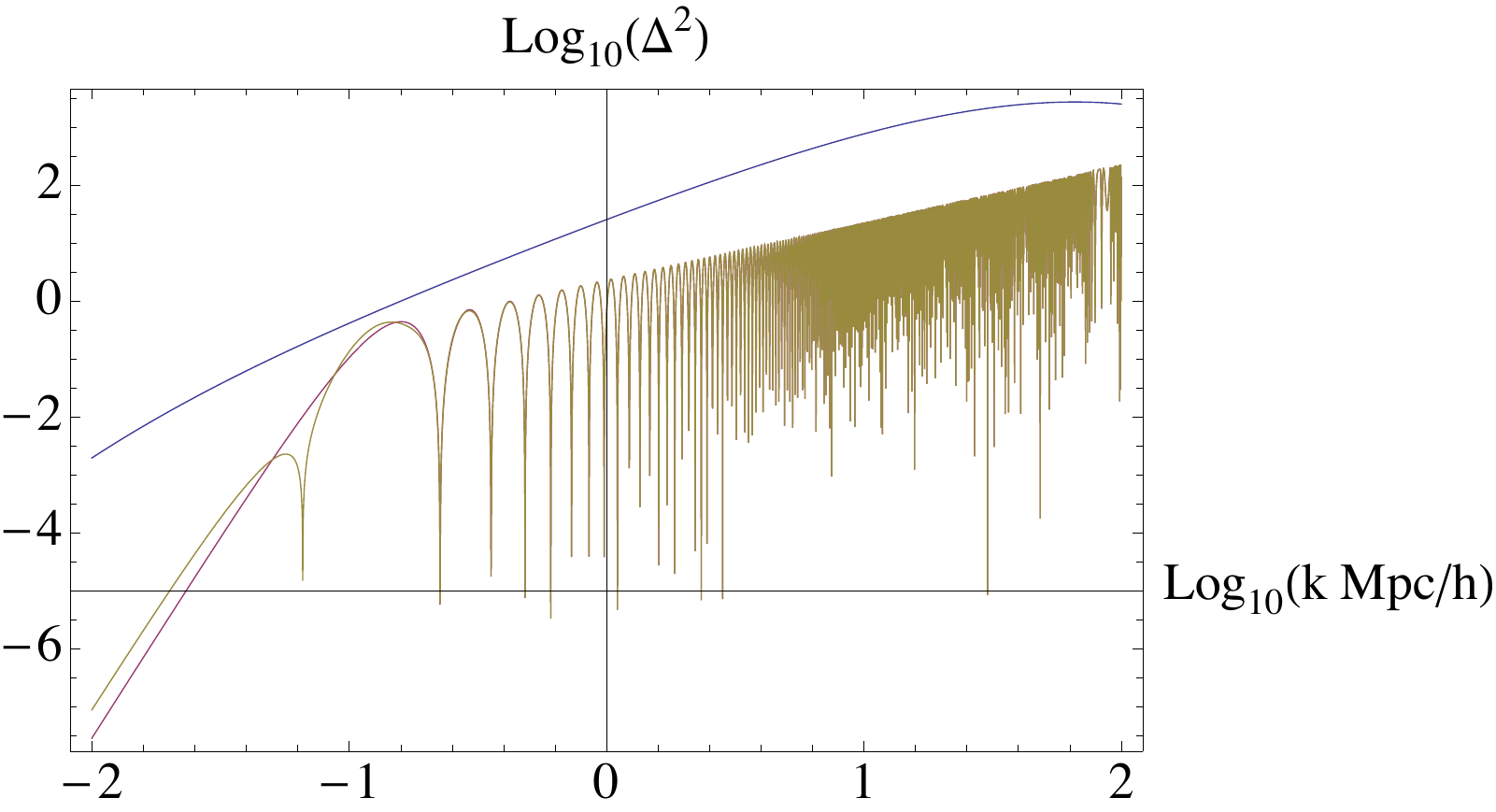}
\fi
\caption{The estimate (\protect{\ref{Delta2US}}) of the matter power spectrum $\Delta(k,z)^2$
for our void distribution, as a function of comoving wavenumber $k$,
evaluated today at $z=0$.
The lower curve includes the correlation correction factor in square
brackets in Eq.\ (\protect{\ref{Delta2US}}), and the middle curve omits it.  The
upper curve is an approximate version of the nonlinear matter power spectrum
at $z=0$ obtained from the Millennium $\Lambda$CDM $N$-body simulation
\cite{Millennium}, shown for comparison.
The parameter values chosen were $H_0 = 73 {\rm km} \, {\rm s}^{-1} \,
{\rm Mpc}^{-1}$, $\Omega_M = 0.3$, $f_0 = f(0) = 0.9$, $R = 35 \, {\rm Mpc}$.}
\label{fig:spectra}
\end{figure}

The power spectrum (\ref{Delta2US}) is shown in Fig.\ \ref{fig:spectra}, both with and
without the correction factor in square brackets that arises from the
correlation between void locations.
For comparison, we also show in
Fig.\ \ref{fig:spectra} an estimate of the nonlinear power
spectrum\footnote{We use
  the following fit to the Millennium power spectrum, obtained from
  Fig.\ 9 of Ref.\ \protect{\cite{Millennium}}: $\Delta(k,z)^2 = \alpha(k)
  (1+z)^{\beta(k)}$, where the functions $\alpha$ and $\beta$ are chosen
  so that $\Delta(k)^2 = 1.40889 + 1.67105 x - 0.11816 x^2 - 0.0356049 x^3 -
0.0367596 x^4$ at $z=0$ and
$\Delta(k)^2 = 0.87558 + 1.56132 x - 0.117482 x^2 - 0.0299214 x^3 -
0.0383988 x^4$ at $z=0.98$, where $x = \log_{10}(k \, {\rm Mpc}/h)$.
This fit is accurate to $\sim 30\%$.} obtained from
the Millennium simulation \cite{Millennium}.
The figure shows that our assumed void model is in rough
agreement with the simulation: the two power spectra agree to within
a factor $\sim 2-3$ at large scales, for $3 \, {\rm Mpc} \alt k^{-1} \alt
30 \, {\rm Mpc}$, but disagree at small scales $k^{-1} \ll 1 {\rm
  Mpc}$, where the Millennium spectrum contains more power.  This is
as expected because our model does not attempt to model structure on these
small scales.

\begin{figure}
\ifx\dofigures\undefined
\else
\includegraphics[scale=0.5]{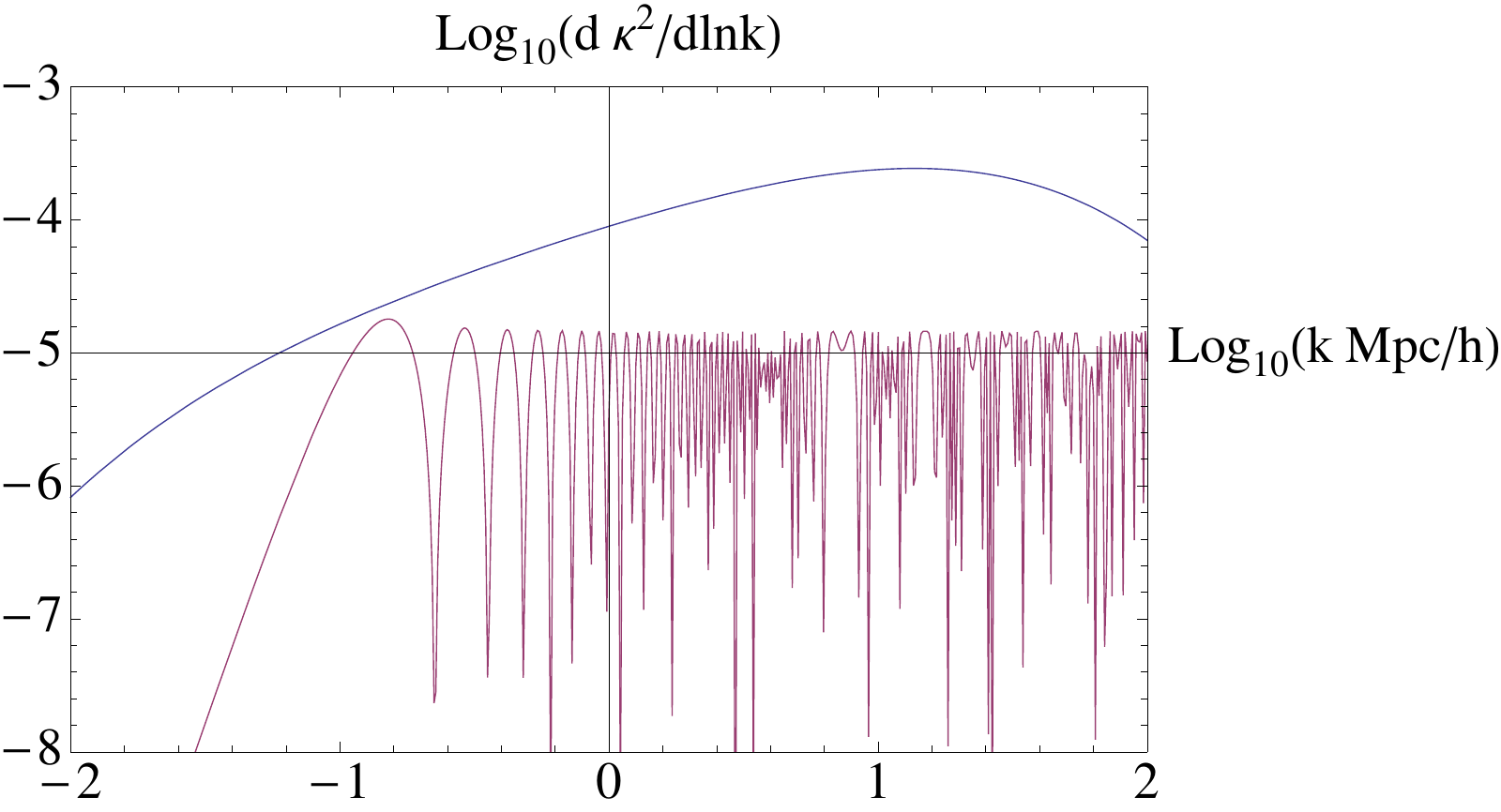}
\fi
\caption{The variance of the lensing convergence per unit
  logarithmic wavenumber, $d \left< \kappa^2 \right>/d\ln k$, for a
  source at redshift $z_s=1$, computed from the spectra shown in Fig.\
  \protect{\ref{fig:spectra}}.  The upper curve is the Millennium
simulation, the lower curve is our void model.}
\label{fig:spectra1}
\end{figure}

We now turn to computing the effects of lensing using these power
spectra.  From the formula (\ref{eq:42-1}) for lensing convergence
$\kappa$ in weak lensing
theory, it follows that for subhorizon modes the variance in
$\kappa$ is \cite{lensing1,lensing2}
\begin{equation}
\left<\kappa^2\right>
=\int d\ln k \left[ \frac{9\pi}{4}H_{0}^{2}\Omega_{M}^{2}\int_{0}^{x_{s}}dx\,w(x,
  x_{s})^2 \frac{\Delta(k,z)^2}{k} \right],
\label{eq:A1}
\end{equation}
where
$x$ is comoving coordinate, $x_{s}$
is the position of the source and
$w=\left(1+z\right)H_{0}x\left(x_{s}-x\right)/x_{s}$ is the lensing
efficiency factor.
The corresponding standard deviation in magnitude shift $\Delta m$ is
$\sigma_m = 5 \sqrt{ \left< \kappa^2 \right> } / \ln 10$, from Eq.\
(\ref{eq:25c}).
We compute the integrand of the $\ln k$ integral
by numerically integrating over redshift,
for a source redshift of $z_s = 1$.
The result is shown in Fig.\ \ref{fig:spectra1}.

Consider first the result for our void distribution.  Fig.\
\ref{fig:spectra1} shows that the envelope of $d \left< \kappa^2
\right>/d\ln k$
asymptotes to a constant at large $k$, indicating a logarithmic
divergence in the variance $\left< \kappa^2 \right>$.
As discussed in the body of the paper, this divergence is an artifact
of our use of distributional density profile for each void, with a
$\delta$-function on the void's surface.  The
divergence can be regulated by endowing
each shell with some small finite thickness $\Delta r$, which is
approximately equivalent to truncating the integral over $k$ in Eq.\
(\ref{eq:A1}) at $k \sim 1/\Delta r$.
Integrating Eq.\ (\ref{eq:A1}) between $10^{-2} {\rm Mpc}^{-1}$ and
$10^2 {\rm
Mpc}^{-1}$ gives the result $\sigma_m = 0.011$, which is substantially smaller
than the result $\sigma_m = 0.031$ obtained from our nonlinear method
in Sec.\ \ref{sec:results} above.  The agreement is improved if we
integrate up to $10^5 {\rm Mpc}^{-1}$, corresponding the effective
cutoff lengthscale in our simulations estimated in
Sec.\ \ref{sec:analyticalmodel} (even though this shell thickness
lengthscale is unrealistic).  In this case $\sigma_m = 0.016$, a
factor of $\sim 2$ smaller than our simulations.
The factor $\sim 2$ disagreement is not too surprising, since
as mentioned above the derivation of
Eq.\ (\ref{eq:A1}) requires the assumption that the density perturbation
is a homogeneous isotropic random process, which is violated to some extent by
our void model.


It is also of interest to compute the standard deviation $\sigma_m$ for the
Millennium simulation spectrum.  Figure \ref{fig:spectra1}
shows that the variance of the lensing convergence per unit
logarithmic
wavenumber $d\left< \kappa^2 \right> / d\ln k$ peaks at
$k \sim 100 \, {\rm kpc}$ (in agreement with Sec.\ 10.5 of Ref.\ \cite{Munshi}).
This indicates that lensing is dominated
by galactic scale structures, as claimed by Holz \& Wald \cite{HW}.
The total standard deviation\footnote{This total standard deviation due to lensing
computed using weak lensing theory and the Millennium simulation agrees
well with that computed
using other methods.  For example, the corresponding standard deviation for $z_s
= 1.5$ is $\sigma_m = 0.066$, which agrees within $\sim 20\%$ with the
standard deviation of the distribution shown in Fig.\ 1 of Ref.\
\protect{\cite{HL}}.} from all scales $10^{-2} {\rm Mpc} \le k^{-1}
\le 10^3 {\rm Mpc}$ is $\sigma_m = 0.044$.  The standard deviation
from integrating only over the scales of voids
$3 {\rm Mpc} \le k^{-1} \le 10^3 {\rm Mpc}$ is $\sigma_m = 0.010$, a
factor $\sim 4$ smaller; this standard deviation agrees well
with our estimate (\ref{finitewidth}) for the thick-wall void model.

\section{Derivation of procedure for computing magnification distribution}
\label{appB}

In this appendix we describe in more detail the derivation of
our prescription for computing magnifications along a ray given by
Eqs.\ (\ref{eq:14}) -- (\ref{eq:18}).

Consider an observer ${\cal O}$ and a source ${\cal S}$.  The angular diameter distance $D_A({\cal O},{\cal S})$ is defined by
\begin{equation}
{D_A^2} = \delta A / \delta \Omega
\label{DAdef}
\end{equation}
where $\delta A$ is the proper area of the source, orthogonal to the
direction to the observer, and $\delta \Omega$ is the observed solid
angle at the observer subtended by the source.  Under a conformal
transformation of the metric, $\delta \Omega$ is invariant while
$\delta A$ transforms by a factor of the conformal factor evaluated at
the source.  It follows that if we define ${\bar D}_A$ to be the
angular diameter distance computed in the conformally transformed
spacetime (\ref{eq:15}), then we have $D_A = a({\cal S}) {\bar D}_A$,
where $a$ is the scale factor.  We now define the magnification
relative to FRW to be\footnote{This definition could equivalently be
  expressed in terms of luminosity distances $D_L$, since $D_L =
  (1+z)^2 D_A$ for any spacetime.}
\begin{equation}
\mu = \frac{D_{A,0}^2} {D_A^2},
\label{mu0}
\end{equation}
where $D_{A,0}$ is the angular diameter distance computed in the
unperturbed FRW model.
Expressing the two angular diameter distances in Eq.\ (\ref{mu0})
in terms of the
conformally transformed versions, the factors of $a({\cal S})$
cancel\footnote{We neglect the contribution to $\mu$ caused by the
  perturbation in the observed redshift of the source, which enters
  when we express the magnification in terms of the observed
  redshift.  This effect gives a subdominant contribution to $\mu$ for
  subhorizon modes \protect{\cite{Bonvin,FRW}}.}, and we obtain
that
\be
\mu = {\bar D}_{A,0}^2 / {\bar D}_A^2 = x_s^2 / {\bar D}_A^2,
\label{muf}
\ee
where $x_s$ is the comoving coordinate of the source.

To compute the angular diameter distance ${\bar D}_A({\cal O},{\cal
  S})$ in the conformally transformed spacetime (\ref{eq:15}), we use
the same method that Holz \& Wald \cite{HW} used in the physical spacetime,
whose derivation we now outline in the context of an arbitrary
spacetime.
Let ${\vec k} = d/dx$ be the past-directed tangent vector to the null
geodesic joining ${\cal O}$ and ${\cal S}$, where $x$ is affine
parameter with $x=0$ at ${\cal O}$.  We choose vectors ${\vec l},
{\vec e}_1, {\vec e}_2$ at
${\cal O}$ so that ${\vec e}_{\hat \alpha} = ({\vec k}, {\vec l},
{\vec e}_A)$, $A = 1,2$ form an orthonormal basis, i.e., satisfy ${\vec
  k}^2 = {\vec l}^2 = {\vec k}\cdot {\vec e}_A = {\vec l} \cdot {\vec
  e}_A  =0$, ${\vec k} \cdot {\vec l} = -1$, ${\vec e}_A \cdot {\vec
  e}_B = \delta_{AB}$.  This orthonormal basis is extended along the
geodesic by parallel transport.

Now let ${\vec \eta}(x)$ be an infinitesimal connecting vector that joins the
geodesic to some nearby geodesic.  The components of ${\vec \eta}$ on
the orthonormal basis satisfy the geodesic deviation equation $d^2
\eta^{\hat \alpha}/dx^2 = - R^{{\hat \alpha}{\hat \beta}{\hat
    \gamma}{\hat \delta}} k_{\hat \beta} k_{\hat \delta} \eta_{\hat \gamma}$.
More explicitly, expanding ${\vec \eta} = \mu {\vec k} + \nu {\vec l}
+ \eta^A {\vec e}_A$, the geodesic deviation equation becomes
\bes
\bea
\label{ddotnu}
{\ddot \nu} &=& 0,  \\
{\ddot \mu} &=& \nu {\cal R} - \eta_C {\cal R}^C , \\
{\ddot \eta}^A &=& \nu {\cal R}^A - \eta_C {\cal R}^{AC}.
\label{gd1}
\eea
\ees
Here dots denote derivatives with respect to $x$, ${\cal R} = R_{abcd}
k^a l^b k^c l^d$, ${\cal R}_A = -R_{abcd} k^a l^b k^c e_A^d$, and
${\cal R}_{AB} = R_{abcd} k^a e_A^b k^c e_B^d$.

We are interested in a set ${\cal B}$ of rays all of which pass through
${\cal O}$ and which define an element of solid angle $\delta \Omega$
at ${\cal O}$.  The corresponding deviation vectors ${\vec \eta}(0)$ must
vanish at ${\cal O}$, and the initial derivatives $d {\vec
  \eta}/dx(0)$ are orthogonal both to ${\vec k}$ and to the four
velocity of the observer, ${\vec u}_{\cal O}$.
If we specialize the choice of orthonormal basis so that ${\vec
  u}_{\cal O} \cdot {\vec e}_A =0$, then it follows that $\nu = {\dot
  \nu} =0$ at ${\cal O}$, and from Eq.\ (\ref{ddotnu}) we obtain that
$\nu(x) =0$ everywhere.  By the linearity of the geodesic deviation equation
it now follows that
\be
\eta^A(x) = {\cal A}^A_{\ B}(x) {\dot \eta}^B(0)
\label{calAdef}
\ee
for some $2\times 2$ matrix ${\cal A}^A_{\ B}$.  This matrix satisfies
the differential equation
(\ref{eq:16}) and initial conditions (\ref{eq:17}) given in Sec.\
\ref{sec:method} above, from Eq.\ (\ref{gd1}) with $\nu=0$.
We define the quantity
\be
\Delta({\cal O},{\cal S}) = \frac{x_s^{2}}{\det \boldsymbol{{\cal
      A}}(x_s)},
\label{VV}
\end{equation}
which is the so-called van Vleck determinant \cite{Visser}.
One can show that this is invariant under rescaling of affine
parameter, under changes of the orthonormal basis that
preserve ${\vec k}$, and under
interchange of ${\cal O}$ and ${\cal S}$.

We now define a set of angular coordinates $\bftheta = \theta^A$ that
parameterize the solid angle measured by the observer, by $\theta^A =
\theta^A_0 + {\dot \eta}^A(0)/({\vec k} \cdot {\vec u}_{\cal O})$,
where $\bftheta_0$ is the direction to the source.  The element of
solid angle is then
\bea
\delta \Omega &=& \int_{\cal B} d^2 \bftheta =
\frac{1}{( {\vec k} \cdot {\vec u}_{{\cal O}})^2}
\int_{\cal B} d^2 {\dot \eta}^A(0) \nn \\
&& =
\frac{1}{( {\vec k} \cdot {\vec u}_{{\cal O}})^2
\ |{\rm det} \boldsymbol{{\cal A}}(x_s)| }
\int_{\cal B} d^2 {\eta}^A(x_s),
\label{deltaOm}
\eea
where we have rewritten the integral using the Jacobian of the
transformation (\ref{calAdef}).

Now consider the element of area $\delta A$ measured at the source
${\cal S}$.
This is defined to be the area in the rest frame of the source,
orthogonal to the direction to the observer.  We choose an orthonormal
basis ${\vec k}, {\vec l}', {\vec e}_A'$ at ${\cal S}$ so that the
four velocity is $({\vec k} + {\vec l}')/2$, and decompose
the connecting vector as ${\vec \eta} = \mu' {\vec k} + \nu' {\vec l}'
+ {\eta'}^{A} {\vec e}_A'$.  Then the area is just $\delta A =
\int_{\cal B} d^2 {\eta'}^A$.  Now the two orthonormal bases
$({\vec k}, {\vec l}, {\vec e}_A)$ and
$({\vec k}, {\vec l}', {\vec e}_A')$
at ${\cal
  S}$ are related by some fixed Lorentz transformation, so we obtain
\bea
\nu &=& \nu', \nn \\
\mu &=& \mu' + \frac{1}{2} \nu' {\bf D}^2 + H_{AB} {\eta'}^A D^B, \nn
\\
\eta^B &=& H_A^{\ B} {\eta'}^A + \nu' D^B,
\eea
for some $SO(2)$ matrix $H_{AB}$ and vector $D^A$.  Since $\nu=0$
everywhere it follows that $\eta^A$ and ${\eta'}^A$ are related by an
$SO(2)$ transformation, which preserves area, and so $\delta A =
\int_{\cal B} d^2 \eta^A(x_s)$.  Combining this with Eqs.\ (\ref{DAdef}),
(\ref{VV}) and (\ref{deltaOm}) now gives for the angular diameter distance
\be
D_A({\cal O},{\cal S})^2 = \frac{x_s^2 ( {\vec k} \cdot {\vec u}_{\cal
    O})^2 }{ |\Delta({\cal O},{\cal S})| }.
\label{DAans}
\ee
This is independent of the normalization of the affine parameter and
of the four-velocity of the source, but does depend on the
four-velocity of the observer.

We now apply the formula (\ref{DAans}) to a stationary observer in the
perturbed Minkowski spacetime (\ref{eq:15}), to obtain the angular
diameter distance ${\bar D}_A$ of Eq.\ (\ref{muf}) above.
Specializing the affine parameter $x$ to be
the comoving coordinate gives ${\vec k} \cdot {\vec u}_{\cal O} =1$,
and then combining Eqs.\ (\ref{muf}), (\ref{VV}) and (\ref{DAans})
gives the magnification formula (\ref{eq:18}).

Finally, we note that in computing the matrix $\boldsymbol{{\cal
    A}}(x_s)$, we follow Holz
\& Wald \cite{HW} in neglecting the influence of the metric perturbation on the
background geodesic, and on the parallel transport of the orthonormal
basis.  The corresponding corrections to the angular diameter distance
have been computed in the weak lensing limit in Refs.\
\cite{Bonvin,FRW} and are subdominant for subhorizon modes, that is, are suppressed by a factor of $(H_0 R)^2$.

\section{Comparison with other studies of lensing due to voids}
\label{appC}

Luminosity distance in the context of Swiss Cheese cosmology has been
studied by Clifton \& Zuntz \cite{CZ}, Brouzakis, Tetradis \& Tzavara
\cite{BTT07,BTT08}, Szybka \cite{Szybka}, Valkenburg \cite{Valkenburg1,Valkenburg2} and Biswas \& Notari
\cite{Biswas}. Other
studies in perturbed FRW cosmologies have been done by Holz \& Wald
\cite{HW} and Hui \& Greene \cite{HG}. In this appendix we summarize
the relevant results
from this literature and compare with our results.

In Clifton \& Zuntz \cite{CZ}, the mean and standard deviation of
apparent
magnitude shifts are studied for redshifts up to $z_s\sim1$ in $\Lambda$CDM
cosmology.
One difference between their study and ours is that
they model voids using a fully relativistic Lemaitre-Tolman-Bondi
model with a smooth choice of density profile,
whereas we use a simpler Newtonian
model where each void
consists of a central uniformly underdense region surrounded by a
zero thickness shell.
Fractional corrections to the Newtonian
approximation scale as $\left(H_{0}R\right)^{2}\sim0.0001$
for 35 Mpc voids, so a fully relativistic void model is not really
necessary; our model is substantially simpler than theirs.
A second difference between the two studies is that
they choose a configuration of voids where the void centers lie along
the line of sight.  Due to this choice, the lensing contributions from successive
voids add coherently instead of random walking, which significantly
changes the magnification probability distribution.
Specifically,
for $z_s=1$ and deep voids,
Clifton \& Zuntz obtain a standard deviation in
modulus shift of $\sigma_m \sim0.01$ (their Fig.\ 16), similar to our value, but they
obtain a mean shift of $\left< \delta m \right> \sim 0.02$, a factor
$\sim 10$ larger than ours.
This difference arises from their lack of randomization of impact
parameters.

Other similar studies are those of Brouzakis, Tetradis \& Tzavara
\cite{BTT08} and Biswas \& Notari \cite{Biswas}.
Brouzakis {\it et al.} also use a fully relativistic Lemaitre-Tolman-Bondi
void model with a smooth choice of density profile.  They find values of
standard deviation $\sigma_m$ which
agree to within $\sim 30\%$ with our model; see their Fig.\ 5 which
applies to $R = 40 $ Mpc voids at $z_s =1$.
Brouzakis {\it et al.} \cite{BTT08} and also Biswas \& Notari
\cite{Biswas} studied the dependence of the magnification distribution
on void sizes, source redshift, and fractional underdensity in the
void interior, and found results which agree qualitatively with ours.
The effects of randomizing void impact parameters
was also studied by Szybka \cite{Szybka}, who found as did we that
the dimming effect due to voids is not enough to mimic the effect of
dark energy.  The
effect of shear is also studied by Szybka, who found its effects to be
very small, in agreement with our results discussed in Sec.\ \ref{sec:nonlin}
above.
The main advantage of our model compared to these studies is simplicity:
our model allows us to explore and understand the effects of a wide
range of parameter values.

Kainulainen \& Marra \cite{KM09,KM11} introduce a
different technique to study lensing. While we compute the probability
distribution of magnifications by doing Monte Carlo simulations of ray
tracing, Kainulainen \& Marra \cite{KM09} develop a method that allows
them to rapidly compute an approximate form of the entire probability
distribution through a combination of numerical and analytical
techniques.  However, their application of this method focus on
the lensing due to galaxies and halos, not on the larger-scale
structures of sheets and voids, so our study is not directly
comparable to theirs.  We note however that it should be possible to
apply their techniques to compute the lensing due to voids.

Finally, a recent paper by Lavallaz \& Fairbairn \cite{Lav} performs a
similar study modeling
voids as 30 Mpc Lemaitre-Tolman-Bondi spheres with Kostov
parameterization \cite{Kostov}. They assume that the supernovae
number density is proportional to the mass density inside voids and
they study the redshift range $0.01 < z < 2.0$. They find that if there
is essentially no cut off in the lower range of $z$, the scatter in
the inferred equation of state parameter $w$ is about $10\%$, while
imposing a cut off in the lower range of $z$
decreases the scatter.

\end{document}